\documentclass[UTF8,a4paper]{article}
\usepackage{amsmath}
\usepackage{graphicx}
\usepackage{subfigure}
\usepackage{epstopdf}
\usepackage{geometry}
\usepackage{ulem}
\usepackage{indentfirst}
\usepackage{amsfonts,amssymb,dsfont}
\usepackage{setspace}
\usepackage{threeparttable}
\usepackage{amsmath,mathtools,amsthm}
\usepackage{array}
\usepackage{extarrows}
\usepackage{upgreek}

\makeatletter
\renewcommand*\env@matrix[1][\arraystretch]{%
  \edef\arraystretch{#1}%
  \hskip -\arraycolsep
  \let\@ifnextchar\new@ifnextchar
  \array{*\c@MaxMatrixCols c}}
\makeatother
\begin{document}

\title{\bf Geometrical structure and the electron transport properties of monolayer and bilayer silicene
near the semimetal-insulator transition point
in tight-binding model}
\author{Chen-Huan Wu
\thanks{chenhuanwu1@gmail.com}
\\Key Laboratory of Atomic $\&$ Molecular Physics and Functional Materials of Gansu Province,
\\College of Physics and Electronic Engineering, Northwest Normal University, Lanzhou 730070, China}


\maketitle
\vspace{-30pt}
\begin{abstract}

\begin{large}


We investigate the electron properties of the monolayer and bilayer silicene which is the honeycomb lattice consist of silicon atoms,
including the optical conductivity and charged impurity scattering,
due to the quasipatricle Dirac-like behaviors near the K-point of silicene.
The spin, valley, sublattice degrees of freedom are taken into consider in the multi-band tight-binding model.
In momentum space, the scattering matrix
which connects the two bare (without interaction) Green's functions 
in the quasiparticle momentum transport process, could be momentum-independent for the single impuirity configuration,
which is similar to the case with small Coulomb coupling in the low-energy Dirac semimetallic system.
While in the zero-frequency limit,
or the frequency-independent case in the strong Coulomb coupling regime, 
the static polarization can be obtained by the random-phase-approximation,
and it's important for the determination of the screened Coulomb scattering by the charge impurity.
The antiferromagnetic order in the silicene is Hubbard-U-dependent, unlike the square lattice which
with the antiferromagnetic ground state,
and provides the premise of the phase transition from the nonmagnetic semimetal phase to the insulator one.
We found that in the absence of electric field, the critical value of the phase transition from semimetal to insulator is 9 eV 
for the clear monolayer silicene, and is 12 eV for the 2nd AA-stacked bilayer silicene or the dirty monolayer silicene
whcih with impurity strength 4 eV.
While the behaviors of 1st AA-stacked bilayer silicene is found similar to the monolayer one.
The in-plane optical conductivity also shows the same results.
\end{large}
\end{abstract}
\begin{large}
\section{Introduction}

Silicene, a topological insulator (TI) together with it's bilayer form or nanoribbon form which have been synthesized experimentally\cite{Feng B},
has very remarkable properties like the graphene\cite{Ezawa M},
and its atom structure was shown in Fig.1(a), and the hexagonal Brillouin zone (BZ) was shown in Fig.1(b)
with its zigzag edge and armchair edge in two directions.
The low-energy dynamics of silicene can be well described by the Dirac-theory.
The silicene is also a $3p$-orbital-based materials with the noncoplanar low-buckled (with a buckle about $0.46$ \AA\ due to the hybridization between the 
$sp^{2}$-binding 
and the $sp^{3}$-binding (which the bond angle is $109.47^\text{o}$) and 
that can be verified by thr Raman spectrum as shown in the Fig.1(e) 
which with the intense peak at 578 cm$^{-1}$ larger than the planar one and the $sp^{3}$-binding one
\cite{Tao L},
and thus approximately forms two surface-effect like the thin ferromagnet matter) lattice structure.
The bulked structure not only breaks the lattice inversion symmetry,
but also induce a exchange splitting between the upper atoms plane and the lower atom plane and thus forms a emission geometry which allows the
optical interband transitions, which for the graphene can happen only upon a FM substrate\cite{Rader O}.
The FM or AFM order can be formed in monolayer silicene by the magnetic proximity effect
that applying both the perpendicular electric field and in-plane FM or AFM field.
Silicene has much stronger intrinsic spin-orbit coupling (SOC) and stronger interlayer interaction compared to the graphene due to its
heavier atom mass and low-bulked structure, respectively.
The bilayer silicene holds both the topological and SC properties by the, e.g., AFM $d_{1}+id_{2}$-pairing,
rather that the $s$-wave one.
There are four kinds of the stacking way for the bilayer silicene with different overlap and buckled-toward directions,
but all with a Bravais lattice unit cell containing four silicon atoms.
Among these four kinds of bilayer silicene, the AB-bt one has been found as the most stable one\cite{Liu F} and thus naturely has the lowest
formation energy as 0.586 eV.
In the Ref.\cite{Liu F}, the AB-bt bilayer silicene has been found that with largest overlap between the lowest conduction band and 
the highest valence band as 300 meV, which shows that it's in a metal phase and thus with a extended Fermi surface.
In this paper we only focus on the two kinds of AA-stacked bilayer silicene (see Fig.1(d) for the side view) 
which has a higher formation energy but lower cohesive energy
compared to the AB-stacked (Bernal) one, and we only consider the vertical hopping (NN) here
unlike for the AB-stacked one.

\section{Geometrical structure of the layered silicene}
We consider the nearest-neighbor (NN) hopping $t$ and the next-nearest-neighbor (NNN) hopping $t'$ of silicene
in this section 
where we imagine a isotropic honeycomb lattice with particle-hole symmetry (PHS) (see the Schematic diagram in Fig.1)
and ignore the diagonal hopping in hexagonal lattice which is qualitatively unimportant.
Then the transfer Hamiltonian $H_{t}$ can be written as 
\begin{equation}
\begin{aligned}
H_{t}=&\begin{pmatrix} H_{AB'}&H_{AA'}\\ H_{A'A}& H_{A'B}\end{pmatrix},\\
H_{AB'}=&\varepsilon_{A}+t'_{AB'}(4{\rm cos}\frac{\sqrt{3}k_{x}a}{2}{\rm cos}\frac{k_{y}a}{2}+2{\rm cos}k_{y}a),\\
H_{AA'}=H^{*}_{A'A}=&t_{AA'}[{\rm exp}(i\frac{\sqrt{3}k_{x}a}{3})+{\rm exp}(i(-\frac{\sqrt{3}}{6}k_{x}a+\frac{k_{y}a}{2}))
+{\rm exp}(i(-\frac{\sqrt{3}k_{x}a}{6}-\frac{k_{y}a}{2}))],\\
H_{A'B}=&\varepsilon_{A'}+t'_{A'B}(4{\rm cos}\frac{\sqrt{3}k_{x}a}{2}{\rm cos}\frac{k_{y}a}{2}+2{\rm cos}k_{y}a),
\end{aligned}
\end{equation}
where $\epsilon_{A}$ is the on-site energy and $a=3.86$ \AA\ is the lattice constant.
The upper and lower band energies are
\begin{equation} 
\begin{aligned}
E_{+}(k_{x},k_{y})=t{\rm cos}\frac{\sqrt{3}k_{x}a}{3}
+2t{\rm cos}\frac{\sqrt{3}k_{x}a}{6}{\rm cos}\frac{k_{y}a}{2}+4t'{\rm cos}\frac{\sqrt{3}k_{x}a}{2}{\rm cos}\frac{k_{y}a}{2}+2t'{\rm cos}k_{y}a,\\
E_{-}(k_{x},k_{y})=t{\rm cos}\frac{\sqrt{3}k_{x}a}{3}-t{\rm cos}\frac{\sqrt{3}k_{x}a}{6}{\rm cos}\frac{k_{y}a}{2},
\end{aligned}
\end{equation}
The plot of upper band energy is shown in the upper panel of Fig.2.
It's clearly that the lower and upper band energies are split by $t$ and $t'$,
and indeed it's origin from the hybridization of eigenstates 
with plane-wave states.
Thus we can obtain that the bands width and the flatness are related to the $t$ and $t'$.
Then the on-site energy within the single-particle picture can be obtained as\cite{Hasegawa Y}
\begin{equation} 
\begin{aligned}
\varepsilon^{2}_{k}=&3t^{2}+2t^{2}[{\rm cos}(-\frac{\sqrt{3}}{2}k_{x}a+\frac{1}{2}k_{y}a)
+{\rm cos}(-\frac{\sqrt{3}}{2}k_{x}a-\frac{1}{2}k_{y}a)+{\rm cos}(\frac{\sqrt{3}k_{y}a}{3})]\\
&+6t'^{2}+2t'^{2}[{\rm cos}(k_{y}a)+{\rm cos}(-k_{y}a)+
{\rm cos}(\frac{\sqrt{3}}{2}k_{x}a-\frac{k_{y}a}{2})+{\rm cos}(\frac{\sqrt{3}}{2}k_{x}a+\frac{k_{y}a}{2})\\
&+{\rm cos}(-\frac{\sqrt{3}}{2}k_{x}a-\frac{k_{y}a}{2})+{\rm cos}(-\frac{\sqrt{3}}{2}k_{x}a+\frac{k_{y}a}{2})]\\
&+2t'^{2}[{\rm cos}(\sqrt{3}k_{x}a)+{\rm cos}(-\sqrt{3}k_{x})+
{\rm cos}(-\frac{\sqrt{3}}{2}k_{x}a+\frac{3k_{y}a}{2})+{\rm cos}(-\frac{\sqrt{3}}{2}k_{x}a+\frac{3k_{y}a}{2})\\
&+{\rm cos}(\frac{\sqrt{3}}{2}k_{x}a-\frac{3k_{y}a}{2})+{\rm cos}(\frac{\sqrt{3}}{2}k_{x}a-\frac{3k_{y}a}{2})],
\end{aligned}
\end{equation}
where the $\varepsilon_{k}$ also describe the dispersion of such hopping configuration which contain nine different hopping directions.
The charts of $E_{+}$ and $\varepsilon^{2}_{k}$ with different $t'$ and $E_{-}$ are shown in the top and bottom panel of Fig.2, respectively.
We can see that the spatial fluctuations are is enhanced with the increasing $t'$.
That means the system is become less stable for increasing $t'$
and the SDW is reduced in the mean time.
Base on the process which taking account the above band energies $E_{\pm}$,
we can obtain the two-band model with the two energy bands which intersect the Fermi surface.
Note that here we ignore the bulking distance for simplify the calculation,
the tight-binding results considering the bulking distance are presented in the following.

While for the pairing scattering process which mentioned above, the induced new band energies $E_{\pm}$ will have more complicate form,
but the splitting interval and bandwidth are still related to the $t$ and $t'$.
In this case the effective interaction obtained by themethod of random-phase-approximation (RPA)
which provides an excellent approximation for our tight-binding model has the similar form
\begin{equation} 
\begin{aligned}
U_{{\rm eff}}=\frac{1}{N}\sum_{ab,kk'}\Gamma'^{l_{1}l_{2}}(k,k',\omega)c^{\dag}_{l_{1}}(k)c^{\dag}_{l_{1}}(-k)c_{l_{2}}(-k')c_{l_{2}}(k'),
\end{aligned}
\end{equation}
with the effective interaction vertex $\Gamma'^{ab}$ between two Cooper pairs near the FS
\begin{equation} 
\begin{aligned}
\Gamma'^{ab}(k,k',\omega)={\rm Re}\sum_{ab,kk'}\Gamma^{a_{1}a_{2}}_{a_{3}a_{4}}(k,k',0)\Lambda^{l_{1}*}_{a_{1}}(k)\Lambda^{l_{1}*}_{a_{2}}(-k)\Lambda^{l_{2}}_{b_{1}}(-k')\Lambda^{l_{2}}_{b_{2}}(k').
\end{aligned}
\end{equation}
Here $a_{i},\ i=1,2,3,4$ is the cell index and $l_{i},\ i=1,2,3,4$ is the orbit index.
And here we note that the orbital space and the band space are closely connected in the following computations
using the RPA method for which the different bands with different eigenvalues.

Considering the bulking distance in the freestanding silicene, the nearest-neighbor hopping vectors may become
\begin{equation} 
\begin{aligned}
r_{1}=&(\frac{\sqrt{3}k_{x}}{3},0,\frac{\sqrt{3}k_{z}}{3}{\rm tan}\theta),\\
r_{2}=&(\frac{-\sqrt{3}k_{x}}{6},\frac{k_{y}}{2},\frac{\sqrt{3}k_{z}}{3}{\rm tan}\theta),\\
r_{2}=&(\frac{-\sqrt{3}k_{x}}{6},-\frac{k_{y}}{2},\frac{\sqrt{3}k_{z}}{3}{\rm tan}\theta),
\end{aligned}
\end{equation}
where $\theta\approx 12^\text{o}55'$ is the angle between the Si-Si band with the $x-y$ plane,
and thus it has $\frac{\sqrt{3}k_{z}}{3}{\rm tan}\theta\approx-\frac{k_{z}}{2\sqrt{14}}$.
while the next-nearest-neighbor hopping vectors ${\bf r}'$ are not affected.
The resulting nearest-neighbor dispersion are\cite{Guzmán-Verri G G}
\begin{equation} 
\begin{aligned}
\epsilon_{1}=\begin{pmatrix}[1.5] \{1,1,1\}\\ \{1,\frac{1}{4},\frac{1}{4}\}\\
\{1,-\frac{1}{2},-\frac{1}{2}\}\\
\{0,\frac{\sqrt{3}}{2},-\frac{\sqrt{3}}{2}\}\\
\{0,\frac{3}{4},\frac{3}{4}\}\\
\{0,-\frac{\sqrt{3}}{4},\frac{\sqrt{3}}{4}\}\\
\{0,0,0\}\\
\end{pmatrix}
e^{i{\bf k}\cdot{\bf r}}=
\begin{pmatrix}[1.5] \{1,1,1\}\\ \{1,\frac{1}{4},\frac{1}{4}\}\\
\{1,-\frac{1}{2},-\frac{1}{2}\}\\
\{0,\frac{\sqrt{3}}{2},-\frac{\sqrt{3}}{2}\}\\
\{0,\frac{3}{4},\frac{3}{4}\}\\
\{0,-\frac{\sqrt{3}}{4},\frac{\sqrt{3}}{4}\}\\
\{0,0,0\}\\
\end{pmatrix}
\begin{pmatrix}[1.5]
e^{ikr_{1}}\\
e^{ikr_{2}}\\
e^{ikr_{3}}
\end{pmatrix}
\end{aligned}
\end{equation}
which make up the $sp^{3}s^{*}$ model of silicene consider the $\sigma$-band
thus the valence (Kohn-Luttinger) band are more lower than the $sp^{3}$ one and the $\sigma$ band and $\pi$ band can't be crossing with each other in this case,
e.g., for the planar silicene the $\sigma$ band and $\pi$ band also can't be crossing with each other due to the orbital symmetry
unless there exist the intrinsic SOC.
The next-nearest-neighbor dispersion are\cite{Guzmán-Verri G G}
\begin{equation} 
\begin{aligned}
\epsilon_{2}=
\begin{pmatrix}[1.5] \{1,1,1,1,1,1\}\\ \{1,1,\frac{3}{4},\frac{3}{4},\frac{3}{4},\frac{3}{4}\}\\
\{1,1,\frac{1}{4},\frac{1}{4},\frac{1}{4},\frac{1}{4}\}\\
\{1,-1,-\frac{1}{2},\frac{1}{2},\frac{1}{2},-\frac{1}{2}\}\\
\{0,0,\frac{3}{4},\frac{3}{4},\frac{3}{4},\frac{3}{4}\}\\
\{0,0,-\frac{\sqrt{3}}{4},-\frac{\sqrt{3}}{4},\frac{\sqrt{3}}{4},\frac{\sqrt{3}}{4}\}\\
\{0,0,\frac{\sqrt{3}}{2},-\frac{\sqrt{3}}{2},\frac{\sqrt{3}}{2},-\frac{\sqrt{3}}{2}\}\\
\{0,0,0,0,0,0\}\\
\end{pmatrix}
e^{i{\bf k}\cdot{\bf r}'}=
\begin{pmatrix}[1.5] \{1,1,1,1,1,1\}\\ \{1,1,\frac{3}{4},\frac{3}{4},\frac{3}{4},\frac{3}{4}\}\\
\{1,1,\frac{1}{4},\frac{1}{4},\frac{1}{4},\frac{1}{4}\}\\
\{1,-1,-\frac{1}{2},\frac{1}{2},\frac{1}{2},-\frac{1}{2}\}\\
\{0,0,\frac{3}{4},\frac{3}{4},\frac{3}{4},\frac{3}{4}\}\\
\{0,0,-\frac{\sqrt{3}}{4},-\frac{\sqrt{3}}{4},\frac{\sqrt{3}}{4},\frac{\sqrt{3}}{4}\}\\
\{0,0,\frac{\sqrt{3}}{2},-\frac{\sqrt{3}}{2},\frac{\sqrt{3}}{2},-\frac{\sqrt{3}}{2}\}\\
\{0,0,0,0,0,0\}\\
\end{pmatrix}
\begin{pmatrix}[1.5]
e^{ikr'_{1}}\\
e^{ikr'_{2}}\\
e^{ikr'_{3}}\\
e^{ikr'_{4}}\\
e^{ikr'_{5}}\\
e^{ikr'_{6}}
\end{pmatrix}
\end{aligned}
\end{equation}
as a $sp^{3}$ model which contains the effect of $\pi-\sigma$ rehybridization\cite{Guo Z X}.
It's obviously that the Eq.(1) consider the $\pi$-band (mainly contributed by the $p$-orbit) which contains both the nearest-neighbor hopping 
and the next-nearest-neighbor hopping, and it can be represented by\cite{Guzmán-Verri G G}
\begin{equation} 
\begin{aligned}
H_{\pi}=
\begin{pmatrix}[1.5]
E_{p}+V_{pp\pi}^{\{2\}}e^{i{\bf k}\cdot{\bf r}'}&V_{pp\pi}^{\{1\}}e^{i{\bf k}\cdot{\bf r}}\\
V_{pp\pi}^{\{1\}}(e^{i{\bf k}\cdot{\bf r}})^{*}&\epsilon_{p}+V_{pp\pi}^{\{2\}}e^{i{\bf k}\cdot{\bf r}'}
\end{pmatrix}
\end{aligned}
\end{equation}
where $V_{pp\pi}^{\{1\}}$ and $V_{pp\pi}^{\{2\}}$ are the first-order and second-order parameters of $\pi$ band made by the $p$ bands,
and $E_{p}$ is the $p$ band's energy,
and 
\begin{equation} 
\begin{aligned}
H_{\pi}=
\begin{pmatrix}[1.5]
0&0\\
0&0
\end{pmatrix}
\end{aligned}
\end{equation}
only in the point of $(k_{x}=0,k_{y}=0)$, i.e., the gapless Dirac-point (with the heavy particle/hole subband),
which have the zero effective mass $m^{*}=0$ for the charge carriers.
The total Hamiltonian is (we omitt the $e^{i{\bf k}\cdot{\bf r}}$ and $e^{i{\bf k}\cdot{\bf r}'}$ for simplicity in the following)
\begin{equation} 
\begin{aligned}
H_{\sigma/\pi}&=
\begin{pmatrix}[1.5]
H_{\pi}&N_{2\times 6}\\
N_{6\times 2}^{\dag}&H_{\sigma}
\end{pmatrix},\\
H_{\sigma}=&
\begin{pmatrix}[1.5]
L&T\\
T^{\dag}&L
\end{pmatrix},\\
T=&
\begin{pmatrix}[1.5]
V_{pp\sigma}^{\{1\}}\{1,\frac{1}{4},\frac{1}{4}\}+V_{pp\pi}^{\{1\}}\{0,\frac{3}{4},\frac{3}{4}\}&(V_{pp\sigma}^{\{1\}}-V_{pp\pi}^{\{1\}})\{0,-\frac{\sqrt{3}}{4},\frac{\sqrt{3}}{4}\}&-V_{sp\sigma}^{\{1\}}\{1,-\frac{1}{2},-\frac{1}{2}\}\\
(V_{pp\sigma}^{\{1\}}-V_{pp\pi}^{\{1\}})\{0,-\frac{\sqrt{3}}{4},\frac{\sqrt{3}}{4}\}&V_{pp\sigma}^{\{1\}}\{0,\frac{3}{4},\frac{3}{4}\}+V_{pp\pi}^{\{1\}}\{0,\frac{3}{4},\frac{3}{4}\}&-V_{sp\sigma}^{\{1\}}\{0,\frac{\sqrt{3}}{2},-\frac{\sqrt{3}}{2}\}\\
V_{sp\sigma}^{\{1\}}\{1,-\frac{1}{2},-\frac{1}{2}\}&V_{sp\sigma}^{\{1\}}\{0,\frac{\sqrt{3}}{2},-\frac{\sqrt{3}}{2}\}&V_{pp\pi}^{\{1\}}\{1,1,1\}
\end{pmatrix},\\
L=&
\begin{pmatrix}[1.5]
L_{1}&L_{3}&0\\
L_{3}^{\dag}&L_{2}&0\\
0&0&E_{p}+V_{pp\pi}^{\{2\}}\{1,1,1,1,1,1\}+\Delta_{sp}
\end{pmatrix},\\
L_{1}=&E_{p}+V_{pp\sigma}^{\{2\}}\{0,0,\frac{3}{4},\frac{3}{4},\frac{3}{4},\frac{3}{4}\}+V_{pp\pi}^{\{2\}}\{1,1,\frac{1}{4},\frac{1}{4},\frac{1}{4},\frac{1}{4}\},\\
L_{2}=&E_{p}+V_{pp\sigma}^{\{2\}}\{1,1,\frac{1}{4},\frac{1}{4},\frac{1}{4},\frac{1}{4}\}+V_{pp\pi}^{\{2\}}\{0,0,\frac{3}{4},\frac{3}{4},\frac{3}{4},\frac{3}{4}\},\\
L_{3}=&(V_{pp\sigma}^{\{2\}}-V_{pp\pi}^{\{2\}})\{0,0,-\frac{\sqrt{3}}{4},-\frac{\sqrt{3}}{4},\frac{\sqrt{3}}{4},\frac{\sqrt{3}}{4}\}，
\end{aligned}
\end{equation}
where $\Delta_{sp}$ is the energy difference between the $3s$ and $3p$ orbits,
which is corresponds to the Kana-Mele term as $\frac{\hbar}{m_{0}}\langle s|p\rangle$.
The next-nearest-neighbor $V_{ss\sigma}^{(2)}=0$ and $V_{sp\sigma}^{(2)}=0$\cite{Grosso G},
for the specific parameters, see the Refs.\cite{Min H,Liu C C,Guzmán-Verri G G,Grosso G} and the references therein.
We can also know that the $H_{\sigma}\neq 0$ even in the Dirac-point unlike the $H_{\pi}$,
and don't relay on the effective mass of charge carriers but the rest mass $m_{0}$.
In the following, we use continuum approximation around the Dirac-point as $t\sum_{{\bf r}}e^{i{\bf k}\cdot{\bf r}}+t'\sum_{{\bf r}'}e^{i{\bf k}\cdot{\bf r}'}
=\hbar v_{F}{\bf k}$ in the following,
where $v_{F}=\hbar{\bf k}/(2m)$ for the free electrons.

In the basis of the perturbative ${\bf k}\cdot {\bf p}$ theory which is widely used for the semiconductor system,
with the twofold degenerate dispersion in the $\Gamma-$point which is comtributed by the $p$ orbits,
and with the bare wave function\cite{Liu C X}
\begin{equation} 
\begin{aligned}
H=\frac{\hbar}{m_{0}}{\bf k}\cdot{\bf p}=\frac{\hbar}{m_{0}}{\bf k}\cdot\langle p_{+}|-i\hbar \partial_{{\bf r}}|p'_{-}\rangle,
\end{aligned}
\end{equation}
with the center momentum formed by two electron states $p_{+}$ and $p_{-}$ with distinct angular momentums, and suffer a perturbation ${\bf k}$.
In the case of TRI, the momentum operator has ${\bf p}={\bf p}^{*}$.
For two sublattices in a unit cell, the momentum matrix element ${\bf p}_{ij}=\langle \psi_{A}({\bf k})|{\bf p}|\psi_{B}({\bf k})\rangle$ 
which is not zero since the inversion symmetry is broken,
and is related to the Wannier function as $\psi_{A}({\bf k})=\sum_{A}w({\bf r}-{\bf r}_{A})e^{i{\bf k}\cdot{\bf r}_{A}}$,
$\psi_{B}({\bf k})=\sum_{A}w({\bf r}-{\bf r}_{B})e^{i{\bf k}\cdot{\bf r}_{B}}$.

We represent the total Hamiltonian which under a perturbation (which may be origin from, e.g., a inhomogenerate electric field or electromagnrtic wave) as
\begin{equation} 
\begin{aligned}
H=H_{0}({\bf p})+\delta H(\partial_{{\bf r}})
\end{aligned}
\end{equation}
the perturbation tiled the spin order by a angle $\theta={\bf k}\cdot{\bf r}$ basis on a initial phase factor $\phi$ which defined above.
For the Zeeman field-induced perturbation, we can perfrom the the canonical transformation to the total Hamiltonian as
\begin{equation} 
\begin{aligned}
H\rightarrow e^{H_{M}}He^{-H_{M}},\\
H_{d}=\begin{pmatrix}[1.5] 0&-M_{z}\\
M_{z}^{\dag}&0\\
\end{pmatrix},
\end{aligned}
\end{equation}
and the SOC term $N_{2\times 6}$ has the below relation with the Zeeman effect\cite{Yao Y}
\begin{equation} 
\begin{aligned}
N_{2\times 6}=M_{z}H_{\sigma}-H_{\pi}M_{z},
\end{aligned}
\end{equation}
Then the above matrix element $T$ under the perturbation-induced rotation is 
$T({\bf k},\partial_{{\bf r}})=\sum_{{\bf k}}\mathcal{R}_{z}^{\dag}T({\bf k})\mathcal{R}_{z}e^{i{\bf k}\cdot {\bf r}}$ with the rotation arounds the
$z$-axis as\cite{Farrell A,Yao Y}
\begin{equation} 
\begin{aligned}
R_{z}&=e^{i\theta}=\begin{pmatrix}[1.5]
{\rm cos}\phi&-{\rm sin}\phi&0\\
{\rm sin}\phi&{\rm cos}\phi&0\\
0&0&1
\end{pmatrix},
&\theta^{{\rm odd}}=\begin{pmatrix}[1.5]
0&i&0\\
-i&0&0\\
0&0&0
\end{pmatrix},
&\theta^{{\rm even}}=\begin{pmatrix}[1.5]
1&0&0\\
0&1&0\\
0&0&0
\end{pmatrix},\\
e^{i{\bf k}\cdot {\bf r}}{R}_{z}=&\begin{pmatrix}[1.5]
-{\rm cos}\phi\ {\rm sin}({\bf k}\cdot{\bf r})&-{\rm sin}\phi&{\rm cos}\phi\ {\rm cos}({\bf k}\cdot{\bf r})\\
-{\rm sin}\phi\ {\rm sin}({\bf k}\cdot{\bf r})&{\rm cos}\phi&{\rm sin}\phi\ {\rm cos}({\bf k}\cdot{\bf r})\\
-{\rm cos}({\bf k}\cdot{\bf r})&0&-{\rm sin}({\bf k}\cdot{\bf r})
\end{pmatrix}.\\
\end{aligned}
\end{equation}

\section{Tight-binding model}
Firstly
the four-band tight-binding (TB) model for the monolayer silicene in low-energy and under both the perpendicular electron field and exchange field,
is given in
a non-Hermitian form\cite{Wu C H,Liu C C,Ezawa M22,Ezawa M4,Ezawa M3,Ezawa M,Zhang J} 
\begin{equation} 
\begin{aligned}
H_{monolayer}=&t\sum_{\langle i,j\rangle ;\sigma}c^{\dag}_{i\sigma}c_{j\sigma}
+i\frac{\lambda_{{\rm SOC}}}{3\sqrt{3}}\sum_{\langle\langle i,j\rangle\rangle ;\sigma\sigma'}\upsilon_{ij}c^{\dag}_{i\sigma}\sigma^{z}_{\sigma\sigma'}c_{j\sigma'}
- i\frac{2R}{3}\sum_{\langle\langle i,j\rangle\rangle ;\sigma\sigma'}c^{\dag}_{i\sigma}(\mu\Delta({\bf k}_{ij})\times {\bf e}_{z})_{\sigma\sigma'}c_{i\sigma'}\\
&+iR_{2}(E_{\perp})\sum_{\langle i,j\rangle;\sigma\sigma'}c^{\dag}_{i\sigma}(\Delta({\bf k}_{ij})\times {\bf e}_{z})_{\sigma\sigma'}c_{i\sigma'}
-\frac{\overline{\Delta}}{2}\sum_{i\sigma} c^{\dag}_{i\sigma}\mu E_{\perp}c_{i\sigma}\\
&+M_{s}\sum_{i\sigma}c^{\dag}_{i\sigma}\sigma_{z}c_{i\sigma}
+M_{c}\sum_{i\sigma}c^{\dag}_{i\sigma}c_{i\sigma}
+U\sum_{i}\mu n_{i\uparrow}n_{i\downarrow},
\end{aligned}
\end{equation}
where $t=1.6$ eV is the nearest-neoghbor hopping which contains the contributions from both the $\pi$ band 
and $\sigma$ band.
The gap function is $\Delta({\bf k})={\bf d}({\bf k})\cdot{\pmb \sigma}$ which in a coordinate independent but spin-dependent representation.
The ${\bf k}$-dependent unit vector ${\bf d}({\bf k})$ here has
${\bf d}({\bf k})=[t'_{SOC}{\rm sin}k_{x},t'_{SOC}{\rm sin}k_{y},M_{z}-2B(2-{\rm cos}k_{x}+{\rm cos}k_{y})]$
for the BHZ model,
where $B$ is the BHZ model
-dependent parameter and $M_{z}$ the Zeeman field term which dominate the surface magnetization
but can be ignore when a strong electric field or magnetic field is applied.
$\langle i,j\rangle$ and $\langle\langle i,j\rangle\rangle $ denote the nearest-neighbor (NN) pairs and the next-nearest-neighbor (NNN) pairs, respectively. 
$\mu=\pm 1$ denote the $A$ ($B$) sublattices. 
Here 
${\bf d}({\bf k}_{ij})=\frac{{\bf d}_{ij}}{|{\bf d}{ij}|}$ is the NNN hopping vector.
$\lambda_{SOC}=3.9$ meV is the intrinsic spin-orbit coupling (SOC) strength which is much larger than the monolayer graphene's (0.0065 meV\cite{Guinea F}).
$R$ is the small instrinct Rashba-coupling due to the low-buckled structure, which is related to the helical bands
(helical edge states) and the SDW in silicene, and it's disappear in the Dirac-point ($k_{x}=k_{y}=0$).
$R_{2}(E_{\perp})$ is the extrinsic Rashba-coupling induced by the electric field.
The existence of $R$ breaks U(1) spin conservation (thus the $s^{z}$ is no more conserved) and the mirror symmetry of silicene lattice.
$M=M_{s}+M_{c}$ is the exchange field which breaks the spatial-inverse-symmetry and the $M_{s}$ is related to the out-of-plane FM exchange field
with parallel alignment of exchange magnetization
and $M_{c}$ is related to the CDW,
which endows sublattice pseudospin the $z$-component\cite{Ezawa M6}.
While for the out-of-plane AFM exchange field $M_{s}^{AFM}$ which is not contained here with antiparallel alignment of exchange magnetization.
Here the $M$ is applied perpendicular to the silicene, and it can be rised by proximity coupling to the ferromagnet\cite{Ezawa M}.
Thus the induced exchange magnetization along the $z$-axis between two sublattices-plane is
related to the SOC, Rashba-coupling, and even the Zeeman-field since it will affects the magnetic-order in $z$-direction.
In fact, if without the exchange field and only exist the SOC, the spin-up and spin-down states won't be degenerates but will mixed around the crossing
points between the lowest conduction band and the highest valence band just like the spin-valley-polarized semimetal (SVPSM).
Note that here we follow the definition of semimetal that the conduction band and valence band have a small overlap,
no matter the two bands are with linear dispersion in the crossing point or 
parabolic dispersion (quadratic) in the crossing point like the Fermi point of the AB-stacked bilayer silicene or graphene.
$\upsilon_{ij}=({\bf d}_{i}\times{\bf d}_{j})/|{\bf d}_{i}\times{\bf d}_{j}|=1(-1)$ 
when the next-nearest-neighboring hopping of electron is toward left (right),
with ${\bf d}_{i}\times{\bf d}_{j}=\sqrt{3}/2(-\sqrt{3}/2)$.
The term contains the exchange field $M$ is the staggered potential term induced by the buckled structure
which breaks the particle-hole symmetry.
Here the coordinate-independent representation of the Rashba-coupling terms is due to the broken of inversion symmetry as well as the mirror symmetry.
The last term is the Hubbard term with on-site interaction $U$ which doesn't affects the bulk gap here but affects the edge gap.
Thus the $U$ is setted as zero within the bulk but nonzero in the edge, 
which is also consistent with the STM-result of silicene that the edge states have 
higher electron-density than the bulk.
And here we take account the on-site Hubbard interaction only and ignore the long-range ones which are screened by the finite DOS with high energy,
like the NN or NNN Coulomb repulsion, interlayer Coulomb repulsion,
and even the one with a range much larger that $a$ (like the Bohr radius in semiconductor).
For the bilayer silicene, we consider two kinds of the AA-stacked silicene: one
with the nearest layer distance as $d=$5.2 \AA\ and intra-layer bond length 2.28 \AA\ with the bulked distance $\overline\Delta=0.46$ \AA\ the smae as the monolayer one
and the another one with the nearest layer distance as $d=$2.46 \AA\ and intra-layer bond length 2.32 \AA\ with the lattice constant $a=3.88$
and the bulked distance $\overline\Delta=0.64$ \AA\
as plotted in the Fig.1.
Thus for the bilayer silicene, the eight-band tight-binding (TB) model in low-energy Dirac theory is
\begin{equation} 
\begin{aligned}
H_{bilayer}=&t\sum_{\langle i,j\rangle ,\sigma,l}c^{\dag}_{i\sigma l}c_{j\sigma l}
+i\frac{\lambda_{{\rm SOC}}}{3\sqrt{3}}\sum_{\langle\langle i,j\rangle\rangle ;\sigma\sigma'}\upsilon_{ij}c^{\dag}_{i\sigma l}\sigma^{z}_{\sigma\sigma'}c_{j\sigma' l}
- i\frac{2R}{3}\sum_{\langle\langle i,j\rangle\rangle ,\sigma\sigma',l}c^{\dag}_{i\sigma l}(\mu\Delta({\bf k}_{ij})\times {\bf e}_{z})_{\sigma\sigma' }c_{i\sigma' l}\\
&+iR_{2}(E_{\perp})\sum_{\langle i,j\rangle,\sigma\sigma',l}c^{\dag}_{i\sigma}(\Delta({\bf k}_{ij})\times {\bf e}_{z})_{\sigma\sigma'}c_{i\sigma'l}
-\frac{\overline{\Delta}}{2}\sum_{i\sigma l} c^{\dag}_{i\sigma l}\mu E_{\perp}c_{i\sigma l}\\
&+M_{s}\sum_{i\sigma l}c^{\dag}_{i\sigma l}\sigma_{z}c_{i\sigma l}
+M_{c}\sum_{i\sigma l}c^{\dag}_{i\sigma l}c_{i\sigma l}
+U\sum_{i,l}\mu n_{i,l\uparrow}n_{i,l\downarrow}+t_{1}\sum_{i,\sigma,l}c_{i}^{\dag}c_{j}\\
&+i\lambda_{SOC}^{{\rm int}}\sum_{i\in A_{1},j\in A_{2},\sigma}c^{\dag}_{i\sigma}(\mu\Delta({\bf k}_{ij})\times {\bf e}_{z})_{\sigma\sigma'}c_{i\sigma'}\\
&+i\lambda_{SOC}^{{\rm int}}\sum_{i\in B_{1},j\in B_{2},\sigma}c^{\dag}_{i\sigma}(\mu\Delta({\bf k}_{ij})\times {\bf e}_{z})_{\sigma\sigma'}c_{i\sigma'}\\
&+\left\{
\begin{array}{rcl}
t_{3}\sum_{i\in A_{1},j\in A_{2},\sigma} c^{\dag}_{i\sigma}\mu c_{j\sigma}+t_{2}\sum_{i\in B_{1},j\in B_{2},\sigma} c^{\dag}_{i\sigma}\mu c_{j\sigma},&\ for\ 2nd\ AA-stacked\ bilayer\ silicene,\\
t_{1}\sum_{i\in A_{1},j\in A_{2},\sigma} c^{\dag}_{i\sigma}\mu c_{j\sigma}+t_{1}\sum_{i\in B_{1},j\in B_{2},\sigma} c^{\dag}_{i\sigma}\mu c_{j\sigma},&\ for\ 1st\ AA-stacked\ bilayer\ silicene,
\end{array}
\right.
\end{aligned}
\end{equation}
where $l=\pm 1$ is the layer index, and $\lambda_{SOC}^{{\rm int}}=0.5$ meV is the interlayer SOC\cite{Ezawa M2}.

Note that for low-energy case the energy spectrum is 
$\varepsilon_{\eta}({\bf k})=\sqrt{\hbar^{2}v_{F}^{2}{\bf k}^{2}+m_{D}^{2}}$ where $m_{D}$ is the Dirac mass generated by the bulk-gap-open
through the spontaneous symmetry breaking by applying the optical field, electric field, magnetic field as explored in Ref.\cite{Wu C H},
even without the Zeeman splitting.
$\eta=\pm 1$ is the valley index for the K and K' valley.
While for the higher energy relativistic case, the Dirac-mass term in above energy spectrum expression show be replaced by $mv_{F}^{2}$
with the relativistic mass $m$.

In the above TB model, both the NNN (linear-Rashba) nd NN (electric-field-induced) Rahsba-coupling is considered,
Since there without the contributions from Dresselhaus term,
the spin SU(2) symmetry is broken together with the effect of NNN SOC term $\lambda_{SOC}$,
but the valley SU(2) symmetry may remains.

In the presence of both the $E_{\perp}$ and the first-order and second-order Rashba-coupling,
the system can be described by
$H=\Psi^{\dag}H^{\pm}_{{\rm eff}}\Psi/2$,
the BCS-like effective Hamiltonian of the 
neighbor valleys by the low-energy Dirac theory in the basis of $\{\tau\otimes\sigma\}$ which reflected in the two-component 
spinor-valued field operators as 
$\Psi=[(\psi_{\uparrow}^{A},\psi_{\downarrow}^{A},\psi_{\uparrow}^{B},\psi_{\downarrow}^{B}),((\psi_{\uparrow}^{A\dag},\psi_{\downarrow}^{A\dag},\psi_{\uparrow}^{B\dag},\psi_{\downarrow}^{B\dag}))]^{T}$, are
\begin{equation} 
\begin{aligned}
H^{+}_{{\rm eff}}=&
\begin{pmatrix} \mathcal{H}({\bf k},\sigma_{z})&\Delta({\bf k},\sigma_{y})\\\Delta^{\dag}({\bf k},\sigma_{y})&\mathcal{H}({\bf k},-\sigma_{z}) \end{pmatrix},\\
\mathcal{H}({\bf k},\sigma_{z})=&\lambda_{{\rm SOC}}\sigma_{z}\tau_{z}+ a R(k_{y}\sigma_{x}-k_{x}\sigma_{y}\tau_{z})
+M\tau_{z}\sigma_{z}-\frac{\overline{\Delta}}{2}E_{\perp}\tau_{z}+\frac{R_{2}(E_{\perp})}{2}(\sigma_{y}\tau_{x}-\sigma_{x}\tau_{y}),\\
\Delta({\bf k},\sigma_{y})=&\begin{pmatrix} i\Delta_{A}&0\\0&i\Delta_{B}\end{pmatrix},\\
\Delta_{A}=&k_{y}\sigma_{y}-ik_{x}\sigma_{x},\\
\Delta_{B}=&-k_{y}\sigma_{y}-ik_{x}\sigma_{x},
\end{aligned}
\end{equation}
and 
\begin{equation} 
\begin{aligned}
H^{-}_{{\rm eff}}=&
\begin{pmatrix} \mathcal{H}({\bf k},-\sigma_{z})&-\Delta({\bf k},-\sigma_{y})\\-\Delta^{\dag}({\bf k},-\sigma_{y})&\mathcal{H}({\bf k},\sigma_{z}) \end{pmatrix},
\end{aligned}
\end{equation}
$\Delta_{A}$ and $\Delta_{B}$ are the pairing gaps of two sublattices.
In the case for valley-polarized metal phase (i.e., the SDC state) which is achieveble under the effect of both the 
vertical electric field\cite{Ezawa M} or magnetic field\cite{Xiao D} and the exchange magnetization
especially under the such a strong SOC which will further intensifys the particle-hole asymmetry between the two valleys.
with the broken sublattice-pseudospin symmetry but remain the chiral symmetry between two valleys,
and the valley-hybridization-term $\Delta({\bf k},\sigma_{y})$ in $H^{+}_{{\rm eff}}$ can be replaced by
\begin{equation} 
\begin{aligned}
\mathcal{V}({\bf k})=&\begin{pmatrix} \sqrt{\mathcal{V}_{1}}&-\sqrt{\mathcal{V}_{2}}(k_{x}+ik_{y})\\-\sqrt{\mathcal{V}_{2}}(k_{x}-ik_{y}) &\sqrt{\mathcal{V}_{1}} \end{pmatrix},\\
\end{aligned}
\end{equation}
which is proportional to the exchange effect between two sublattices (or the potential differentce
between two sublattices) with $\mathcal{V}_{1}$ the hybridization gap which is also proportional to the potential differentce 
and $\mathcal{V}_{2}$ the parabolic band dispersion which with a opened gap.
The low-energy effective Hamiltonian can be written as\cite{Ezawa M4,Ezawa M3}
\begin{equation} 
\begin{aligned}
H=\eta\hbar v_{F}(\tau_{x}k_{x}+\tau_{y}k_{y})+\eta\lambda_{{\rm SOC}}\tau_{z}\sigma_{z}+aR\eta\tau_{z}(k_{y}\sigma_{x}-k_{x}\sigma_{y})
-\frac{\Delta}{2}E_{\perp}\tau_{z}+\frac{R_{2}(E_{\perp})}{2}(\eta\sigma_{y}\tau_{x}-\sigma_{x}\tau_{y})+M\tau_{z}\sigma_{z},
\end{aligned}
\end{equation}
and with the eigenvalue
\begin{equation} 
\begin{aligned}
\varepsilon_{\eta}({\bf k})=Ms_{z}\pm\sqrt{\hbar^{2}v_{F}^{2}{\bf k}^{2}+(\frac{\overline{\Delta}}{2}E_{z}+\eta\hbar m_{{\rm w}}-\eta s_{z}\lambda_{{\rm SOC}})^{2}}.
\end{aligned}
\end{equation}
For the bilayer silicene, the BCS-like effective Hamiltonian can be expressed as 
\begin{equation} 
\begin{aligned}
H^{{\rm bilayer}}_{{\rm eff}}({\bf k})=
&\begin{pmatrix} H^{+}_{{\rm eff}} &t_{{\rm inter}}\\t_{{\rm inter}}&H^{-}_{{\rm eff}} \end{pmatrix}.
\end{aligned}
\end{equation}
and the corresponding eigenvalue (band dispersion) is\cite{Nicol E J}
\begin{equation} 
\begin{aligned}
\varepsilon'_{\eta}({\bf k})=\pm\sqrt{m_{D}^{2}+(\frac{t_{{\rm inter}}}{\sqrt{2}})^{2}+\hbar^{2}v_{F}^{2}{\bf k}^{2}+l
\sqrt{(\frac{t_{{\rm inter}}}{\sqrt{2}})^{4}+\hbar^{2}v_{F}^{2}{\bf k}^{2}\cdot(t^{2}_{{\rm inter}}+(2m_{D})^{2})}}.
\end{aligned}
\end{equation}
The bands obtained by this eigenvalue are shifted upward by the positive $t_{{\rm inter}}$ and downward by the negative $t_{{\rm inter}}$, 
which corresponds to the antibonding and bonding states, respectively.
And the energy of antibonding states are incresed with the increasing scattering strength,
and reaches the maximum in the critical point as we detect below.
The interaction terms, like the strong intrinsic SOC are contained in the Dirac mass term of the above expression,
and it also valid for the gap-anisotropy case,
which increse when the long-range Hubbard interactions are taken into consider,
or the long-range hopping case, like the NNN hopping.

The vertex correlation function $\Gamma$ is associated with the jumping of the self-energy
with different spectral weight in momentum space,
and it plays a important role in the variant cluster approximation (VCA).
For the impurities scattering system, the vertex correlation function is arise with the lifting of Fermi level ($E_{F}$),
which leads to the segmental structure of energy spectrum.
and thus increase the cancellation effect\cite{Grimaldi C} to the Hall conductivity.
Otherwise it can be ignored if Fermi energy $E_{F}$ is close to zero with a very low FS and thus it's compeletely spin-dependent.
While for the interband longitudinal conductivity,
it will vanishes when there exists linear $R$ and linear Dresselhaus coupling with equal strength\cite{Li Z}.
Although the vertex function is piecewise for a finite Fermi energy
and in the direction which normal to the boundary between different pieces with different self-energies (or momentum),
it can be ignored for the case of $m_{D}=0,\ t_{{\rm inter}}=0$ through the vertex renormalization\cite{Khveshchenko D V}
and the conductivity can be evaluated by the bubble approximation.

We carry out the first-principle (FP) density functional theory (DFT) calculations 
using the QUANTUM ESPRESSO package\cite{Paolo} with the generalized gradient approximation (GGA),
and the Perdew-Burke-Ernzerhof (PBE)\cite{Perdew J P} exchange correlation is used.
The plane wave energy cutoff is setted as 250 eV in our calculation,
and the structures are relaxed until the Hellmann-Feynman force on the atoms are below 0.01 eV/\AA\ .
As shown in the Fig.3(a), the band structure of monolayer silicene shows linear and isotropic relations $\varepsilon=\pm \sqrt{3}t|{\bf k}-{\bf K}|/2$ 
near the Dirac-point (and also the for DOS) which suggest a
energy-independent group velocity $v_{g}$ even it's slightly gapped up, while the bilayer silicene dose not.
The pseudogap in the such point suggrests the semi-metal phase of silicene which can be charged to other phase 
through the tunable phase transitions in the nanodisk, nanotube, or nanoribbon silicene\cite{Wu C H}.
While the linear dispersion in $M$-point is exist only in the small-$a$ limit.
Note that here we consider the case that the Fermi level lies within in the band gap (the lowest conduction band and the highest valence band)
and at the point with $\mu=0$.
The DOS of monolayer silicene and AA-stacked bilayer silicene are presented in the right-side of the Fig.3.
We found that the PDOS of 1st AA-stacked bilayer silicene almost has the same shape with the monolayers',
but just twice in the amount.
For the 2nd AA-stacked bilayer silicene (Fig.3(c)), which remain the semimetal but the two linear-crossing Dirac-point drift away from the K-point,
and one lies above the Fermi level while the other one lies below the Fermi level which is different from the bilayer graphene.
The total DOS of 2nd AA-stacked bilayer silicene is not zero at Fermi level similar to the AB-stacked graphene,
and exhibit great difference with the monolayer one and the 1st AA-stacked bilayer one.
Thus we only explore the DOS of the monolayer silicene and 2nd AA-stacked bilayer silicene in the following.
The presented band structure of AA-stacked bilayer silicene shows more strong hybridization between the lowest conduction band and highest valenc band
in low-energy than the AB-stacked one.
For the antiferromagnetic (AFM) silicene, the AFM spin order with the NN sublattice pairing (like the Cooper pairs) symmetry $d_{1}+id_{2}$
with chiral superconductivity (SC) is supported by the strong electron-electron repulsive interaction especially at the singular value of DOS like the graphene
\cite{Wu C H}
and with the phase winding number $2\pi$ around the whole hexagonal lattice (or Brillouin zone) (or 
$4\pi$\cite{Nandkishore R} around the Fermi surface (FS) which is hexagonal at full band filling $v_{f}$ when undoped).
The SC pairing strength (or SC gap) $\Delta_{SC}=\Delta_{0}e^{i\theta}$ where $\theta$ is the SC phase as shown in the Fig.1(c).
But for highly spin-polarized case, like the edge states in the single-Dirac-cone (SDC) phase which with one valley has gapless bulk gap
and has also been found in several kinds of three-dimension TI\cite{Zhang H},
the lifted ferromagnetic (FM) may give rises the NNN FM pairing correlation with the $p$-wave or $f$-wave spin-triplet pairing.

Fig.3(d) shows the single particle DOS for the massless Dirac-fermion in the absence of the impurity and lattice defect.
The linear relation appear in the low-energy region $E\ll t$, which is guaranteed by the high order correction of the dispersion 
in the renormalization group theory\cite{Nandkishore R}.
It exhibit great difference with that for the
spinless noninteracting case (including the interlayer interaction).
The DOS-map is calculated by the RG method with the linear dispersion constricted in the bandwidth in the range $\pi$-band -6.4 eV$\sim$6.4 eV,
and the UV cutoff within this RG procedure is setted much smaller that the bandwidth which is $\Lambda\simeq t=1.6$ eV.

In Fig.4, we show the map plot of the tight-binding energy dispersion and the corresponding DOS for the silicene with particle-hole symmetry
where we ignore the broken of inversion symmetry by the bulked structure and the Rashba-coupling (NNN).
In Fig.4(a), we consider the dispersion in hexagonal BZ with the contribution only from $\pi$-electron, 
$t=V_{pp\pi}^{(1)}$
(which measured as -0.72 in Ref.\cite{Guzmán-Verri G G} and -1.12\cite{Liu C C}):
$\varepsilon_{{\rm hex}}=\pm t\sqrt{1+4{\rm cos}^{2}( k_{x}a/2)+4{\rm cos}( k_{x}a/2){\rm cos}(\sqrt{3} k_{y}a/2)}$.
Since the zero-energy point in DOS corresponds to the spin-degenerate point and the Fermi level is in the place with $\mu=0$,
when the upper band (conduction band) is empty, the lower one is fully filled, 
and then give rise to the maximum spin-polarization with the minimum interaction strength to
minimized the system energy due to the Hund's rule. That's also consist with the zero-energy state in the gapless Dirac-point.
In this case the total spin is becomes maximum as $\hbar N/2$ ideally in the ferromagnetic ground state where $N$ is the number of electrons
which equal to four times of the cell number in the four band model (doubly degenerate).
Such fully spin-polarized pattern also appear in the case of SDC and the spin-valley TI as we disussed in the Ref.\cite{Wu C H},
and it also 
excludes the double occupation at half-filling $d_{hf}=\sum_{i}n_{i\uparrow}n_{i\downarrow}/N$.
and thus makes the method of
dynamical mean-field theory (DMFT) which with the Hartree–Fock term and it 
is efficient to dealing with the nonequilibrium problem with different bandgaps\cite{Wu C H,Wu C H2}
lose efficacy in the single-site problem\cite{Go A}.
What's more, the critical value derived from the DMFT: U=2.23$t$\cite{Jafari S A}, is far away from our result obtained in the follwing text,
which is 9 eV$\approx 5.6t$ for the pure monolayer silicene and $12$ eV$\approx 7.5t$ for the monolayer silicene with impurity.
That's due to the DMFT ignore the quantum fluctuation and the up-spin are independent with the down-spin,
and thus there exist great deviation from the ture results in our model, which with drastic spin and charge fluctuations.
That can also be seem from the large otherness from the semielliptic DOS derived from the DMFT (see Ref.\cite{Wu C H}) to our results.

The in-plane spin texture ($\sigma_{x}$ and $\sigma_{y}$) can be obtained by the
spin expectation values as\cite{Li Z} $s_{x\pm}=\pm\frac{\hbar}{2}\frac{\mathcal{R}}{\sqrt{\mathcal{R}^{2}+\mathcal{R}_{2}^{2}}}$,
$s_{y\pm}=\pm\frac{\hbar}{2}\frac{-\mathcal{R}_{2}}{\sqrt{\mathcal{R}^{2}+\mathcal{R}_{2}^{2}}}$,
with $\mathcal{R}=(a\eta\tau_{z}R k_{y}-\frac{R_{2}(E_{\perp})}{2}\tau_{y})$,
$\mathcal{R}_{2}=(-a\eta\tau_{z}Rk_{x}+\frac{R_{2}(E_{\perp})}{2}\tau_{x})$,
and the $\pm$ here corresponds to the sign of Rashba energy $E=\frac{m(R^{2}+(R_{2}(E_{\perp})^{2}))}{2\hbar^{2}}$.
While the out-of-plane spin texture is related to the Dirac mass and the Zeeman splitting.
The group velocity through the above Dirac Hamiltonian as
$v_{gx}=\frac{\partial \mathcal{H}}{\hbar\partial k_{x}}=v_{F}\eta\tau_{x}-\frac{1}{\hbar}a\eta\tau_{z}R\sigma_{y}$,
$v_{gy}=\frac{\partial \mathcal{H}}{\hbar\partial k_{y}}=v_{F}\tau_{y}+\frac{1}{\hbar}a\eta\tau_{z}R\sigma_{x}$.
The effect of Rashba-coupling to the DOS is exploed in the following section.

\section{DOS}

In the energy-dependent density of state (DOS), the van Hove singularities (VHS) emerge at the band filling $\delta=\pm t$
away from the Dirac-point and with the FS nesting which with infinity 
(impurity) quasiparticle lifetime.
Although the large DOS in the VHS effectively enhance the effect of interaction, the strength of interaction is linearly increase with the increasing 
scattering rate and the logarithmically divergent DOS toward the Dirac-point.
The large DOS in VHS also effectively screen the long-range Coulomb repulsion,
and diverges the susceptibility.
The VHS also exhibit peak signal in the twisted silicene or graphene due to the new saddle point created by the ratotaed Dirac-point
which are still in the K-point\cite{Li Z2}.
The VHS
origin by the saddle points of $\pi$ and $\pi^{*}$ bands in BZ which is the in the $M$-point for unstrained silicene,
are keep away from the K-point to persist the semimetal phase,
and plays a important role in the phase transition to the metal or the band insulator with sizeble bulk gap.
The strong on-site Hubbard interaction (repulsive) also give rise the chiral SC in the bilayer silicene or the graphene\cite{Nandkishore R},
while the long-range Hubbard interaction are screened in this point.
For the case with the lattice defects like the impurities (dopant) or vacancy, the K-point may exhibits a peak of the $\delta$-function
which is obviously smeared by a finite width related to the strength of the impurities or the size of the vacancy,
and makes
the quasiparticle lifetime $\tau$ turn to maximum value (setted as 1) in this case and 
note that the impurities scattering rate which is defined always always $\ge 0$ is $\Gamma=1/(2\tau)\rightarrow \infty$ here.
That's very different from that
in the noninteracting limit as well as the case of Hubbard U=0.
The impurities scattering potential after the Fourier transformation is $V({\bf k}_{s})=\frac{2\pi U}{\sqrt{(\Delta {\bf k})^{2}+{\bf k}_{s}^{2}}}$
with the coulomb potential $U=\frac{qq'}{4\pi\epsilon_{0}\epsilon_{s}}$ where $\epsilon_{0}=1$ is the vacuum dielectric constant
and $\epsilon_{s}=34.33$ is the dielectric constant of silicene.
$\Delta {\bf k}=|{\bf k}-{\bf k}'|=2k\ {\rm sin}\theta$\cite{Adam S,Vargiamidis V}, where $\theta$
describes the difference between the monentums before scattering and after scattering,
and it tends to zero $\theta\rightarrow 0$ for the SC silicene (deposited on a SC electrode or generate the topological superconductor by the STM probe).
The $\Delta {\bf k}$ is zero only for the elastic scattering in which case the scattering potential is close to a $\delta$-function
similar to the Lorentzian representation
and become $\Delta {\bf k}$- and ${\bf k}_{s}$-independent.
In this case, the scattering potential is decay as $1/|{\bf k}_{s}|$.
Due to the exist of the impurities and lattice defects, the quantum spin-Hall effect with the spin-polarized current 
may more observable due to the SOC with the impurities,
even without applying the external exchange field or the electric field,
and it's robust against the nonmagnetic impurity.

At the time-reversal symmetry and particle-hole symmetry points of momentum space\cite{Wu C H}, 
the nonchiral umklapp backscattering term (not contains the forward scattering exchange $J_{z}$) which is\cite{Wu C2}
\begin{equation} 
\begin{aligned}
J_{z}a\int dx e^{-i\phi}\psi^{\dag}_{L\uparrow}(x)\psi^{\dag}_{L\uparrow}(x+a)\times
e^{-i\phi}\psi^{\dag}_{R\downarrow}(x)\psi^{\dag}_{R\downarrow}(x+a)+h.c.
\end{aligned}
\end{equation}
in Fermion language
is allowed, where $\phi=\varphi x$ with the left move and right move N\'{e}el order $\varphi=\pi$ 
(half of the phase of Wigner-Seitz unit cells) and with the scaling dimension 
just be one Luttinger parameter $K$\cite{Zheng D} at commensurate filling\cite{Wu C2}
where the phase transition to a insulator with gap happen. $g$ is the scattering strength factor. 
And here the renormalized Fermi velocity has $v_{F}=1+\frac{J_{z}a}{\pi\hbar v_{F}}(1-{\rm cos}(2k_{F}a))$.
While the chiral term renormalize the Fermi velocity in the homogenerate system without the domain wall,
the inhomogenate case will be discuss below.
For umklapp scattering, $\Delta {\bf k}$ provides a estimation for the momentum transfer on the semimetal Dirca-sea,
and it's proportional to the change of DOS compared to the Fermi momentum $k_{F}$ before scattering.

In low-temperture, since the elastic scattering is dominate, we can simplify the scattering potential as a $\delta$-function
which is momentum-independent with ${\bf k}\approx{\bf k}'$ thus $\theta\approx 0$.
Then without spin-degenerate, the longitudunal in-plane conductivity (diagonal) in linear response theory is\cite{Charbonneau M}
\begin{equation} 
\begin{aligned}
\sigma_{xx}=\sigma_{yy}=\frac{\beta e^{2}}{S}\sum_{m}f_{m}(1-f_{m})\frac{\langle m|v_{x}|m\rangle\langle m|v_{y}|m\rangle}{\omega+i\delta+2\Gamma}
\end{aligned}
\end{equation}
where $S=3\sqrt{3}/2$ is the area of unit cell,
$\omega=(2n+1)\pi/\beta$ is the fermionic Matsubara frequency where $\beta$ is the inverse temperature.
$v_{x}=\frac{\partial}{\hbar\partial k_{x}}$ is the velocity operator.
Here $\sigma_{xx}=\sigma_{yy}$ is tenable for the low-temperature in which the elastic scattering is dominate,
and with velocity operators in the matrix-form:
$v_{x}=\begin{pmatrix}
0&v_{F}\\
v_{F}&0
\end{pmatrix}$, $v_{y}=\eta\begin{pmatrix}
0&iv_{F}\\
-iv_{F}&0
\end{pmatrix}$.
While the transverse off-diagonal in-plane conductivity for the nonelastic scattering is
\begin{equation} 
\begin{aligned}
\sigma_{xy}=\frac{i\hbar e^{2}}{S}\sum_{m\neq n}\frac{f_{m}-f_{n}}{(E_{n}-E_{m}+\hbar(\omega+i\delta)+i\Gamma)(E_{m}-E_{n})}\langle m|v_{x}|n\rangle \langle n|v_{y}|m\rangle.
\end{aligned}
\end{equation}
The scattering rate $\Gamma$ here is defined as
\begin{equation} 
\begin{aligned}
\Gamma=\frac{1}{2\tau}=\frac{\pi n}{\hbar}V^{2}.
\end{aligned}
\end{equation}
Here the charged impurity density $n$ is momentum-independent for the single-impurity case.
The $\Gamma$ can be estimated as 0.01$t=0.016$ eV here and note that the effect of SOC is ignored in this scattering process.
If the SOC is taken into consider in the collision process of the impurity scattering as done in the explores of spin-Hall effect\cite{Dyakonov M I}
with the presence of electric currence and the spin currence,
the $\Gamma$ becomes $<\frac{1}{2\tau}$ and thus the DOS may be increased.
In the domination of impurity scattering and with a certain impurity concentration, the 
spin currence can be described by angle $\theta$ in Maxwell theory.
with the spin-palarized currence perpendicular to the applied electric field which exhibit the quantum quantum anomalous Hall (QAH) effect\cite{Wu C H}.

\section{Results and discussion}

The local DOS provides a good estimation for the diagonal longitudinal conductivity which relys on the interband transitions
and in contrast with the intraband transverse conductivity,
and can be expressed as 
\begin{equation} 
\begin{aligned}
D({\bf k},\omega)=\int_{BZ}\frac{d{\bf k}}{(2\pi)^{2}}(f_{m}-f_{n}){\rm Im}(G_{{\bf k}}(E_{m}-E_{n})-G_{{\bf k}}(E_{n}-E_{m})),\ E_{m}<E_{n},
\end{aligned}
\end{equation}
where $f_{m}=1/(e^{\beta(E_{m}-\mu)}+1)$ is the Fermi-Dirac distribusion function with $\mu$ the chemical potential in Fermi level,
which makes it more like a joint DOS (JDOS) but not 
a single particle DOS,
and the lattice Green's function in helicity basis $G_{{\bf k}}(E_{m}-E_{n})=[E_{m}-E_{n}-(\hbar\omega+i2\Gamma)-\mu]^{-1}$
which can be obtained by the retarded form 
analytical continuation as $i\hbar\omega_{l}\rightarrow \hbar(\omega_{l}+i\delta)$\cite{Wu C H} where $\delta=0^{+}$ is a small positive quantity
and it has $\omega+i\delta\rightarrow 0$ in dc-limit.
The Fermi-Dirac distribusion function can be replaced by the Heaviside step function $\theta$ in the zero-temperature limit
e.g., $f=1$ for the electron-like (occupied) band, and $f=0$ for the hole-like (unoccupied) band
and with the Fermi level lies between them.
$E_{m}$ and $E_{n}$ are the energies of two distrinct electron states.
To see the effects of the Rashba-coupling on the DOS with the spin expectation values deduced in above,
we show the contributions from the Rashba-coupling to the positive-DOS, $D_{R}({\bf k},\omega)$, in Fig.6 where only the NN Rashba-coupling is taken into consider and with the 
applied perpendicular electric field ranges from zero to 8 eV.
Fig.6(a) is for the case that Fermi level is lies within the conduction band and valence band (with $\mu=0$),
and (b) is for the case that Fermi level is lies in the conduction band with $\mu=1$.
We see that with the increase of $\mu$,
the $D_{R}({\bf k},\omega)$ is reduced and the starting point of the transverse axis is shifted.
Except that, the sequence of the initial slopes are changed too:
the largest slope corresponds the $E_{\perp}=1$ eV ($R_{2}=0.012$ eV) in the $\mu=0$ case,
while it corresponds the $E_{\perp}=2$ eV ($R_{2}=0.024$ eV) in the $\mu=1$ case.

We also found that for monolayer silicene,
the DOS in zero-temperature-limit (the silicene is most stable now and similar to the results of non-interacting case) 
is simply detemined by the Dirac-mass, with the certain degenerate number (or degrees of freedom) 4,
as shown in the Fig.5(a),
which is
\begin{equation} 
\begin{aligned}
\rho_{T\rightarrow 0}(\varepsilon,\omega)=
\frac{4|\varepsilon|}{2\pi\hbar^{2}v_{F}^{2}}\frac{1}{2}\sum_{\eta=\pm 1}\left[\theta(|2\varepsilon|-2|m_{D}|_{\eta})\right],
\end{aligned}
\end{equation}
and the Dirac-mass here is setted as $m_{D}=|\lambda_{SOC}+M|=1$ eV.
For bilayer silicene in zero-temperature-limit, the DOS is also dependent on the Dirac-mass,
but the width of the minimum DOS-plateau is the same as the monolayer silicenes', i.e., $2|m_{D}|$, but difference from the AB-stacked bilayer one,
and the value of this minimum DOS-plateau is dependents on the band structure in the Dirac-point.
Fig.5(b) shows the diamagnetic susceptibility which is negative in the low-temperature region for the monolayer silicene.
In fact, both the diamagnetic and paramagnetic response which with opposite magnetic moment (i.e., 
diamagnetic moment and paramagnetic moment with the spin carriers along the edge direction carriers the up- and down- spin, respectively)
are coexist in the silicene due to the interactions between the magnetic field and the charge
carriers with spin-up and spin-down, respectively, and they are both increse with the temperature.

In the presence of nonzero impurity scattering angle with a single impurity, the above expression can be rewritten as
\begin{equation} 
\begin{aligned}
D({\bf k},\omega)&=\int_{BZ}\frac{d{\bf k}}{(2\pi)^{2}}(f_{m}-f_{n})\\
&{\rm Im}
[G_{{\bf k}}(E_{m}-E_{n})T(\omega)G_{{\bf k}}(E_{m}-E_{n}+\Delta{\bf k})-G_{{\bf k}}(E_{n}-E_{m})T(\omega)G_{{\bf k}}(E_{n}-E_{m}+\Delta{\bf k})],
\end{aligned}
\end{equation}
The expression of the DOS is distinct from the conductivity since it takes the imaginary part of the lattice Green's function in the
defect configuration.
Here the effect of $T(\omega)$ is similar to the vertex function except that the vertex function is a connection between different frequencies
but with the same momentum while the $T(\omega)$ here is a connection between different momentums which is related to the scattered wave vector
but with the same frequency.
And here the direction of $T(\omega)$ is perpendicular to the boundary between the two distinguish momentum tiles,
which is weighted by the DOS or the spectral function\cite{Haule K,Go A,Xu W}
and is useful to explore the low-energy behaviours of 
optical conductivity and the Hall conductivity.
For the nonmagnetic impurity
(or the weak magnetic ordering impurities like W- or Mo-silicene), 
the $T(\omega)$ has\cite{Wang Q H}
\begin{equation} 
\begin{aligned}
T^{-1}(\omega)=\frac{1}{V_{s}\sigma_{z}}-\int\frac{d^{2}k}{4\pi^{2}}G_{{\bf k}}(E_{m}-E_{n})
\end{aligned}
\end{equation}
where $V_{s}$ is the single scalar scattering with the $z$-direction spin-polarization.

Fig.7(a) shows the effect of Hubbard U to the JDOS of monolayer AFM silicene
with a half-filled impurity band in the middle.
We can see that the sharp of the JDOS-curve have not obviously changes for the Hubbard U$\le 12$,
and the bandwidth is slightly increased.
Up to $U\ge 13$, the Fermi level was lifted up to the conduction band in a large extent with increasing DOS and leading to a enlarged chemical potential,
similar to the effects of the high doping,
and the bandwidth also largely increased to 4 eV in $U=18$ eV which obviously exhibits the band insulator phase.
Thus it's direct that the critical U,
which is insensitivity to the strength of impurities, 
for such a change is around 12 eV (about 7.5$t$) which is slightly smaller than that of the graphene which is 13.29\cite{Jafari S A},
but close to the value predicted by the Brinkman-Rice analysis which is 11.5 for the 2D Hubbard model\cite{Ebrahimkhas M}.
And it's nearly twice as large as the critical value of the phase transition of metal-to-insulator for the 
half-filling $1/r$ Hubbard chain, which is equal to the bandwidth whose absolute value is 6.4 eV for silicene\cite{Wu C H2,Gebhard F}.
That also implys that the critical value of Hubbard U is associated with the dimension like the DOS distribution\cite{Santoro G}.
For FM monolayer silicene with nonzerp spin-polarization, the curves of JDOS shows the similar behavior but just with larger band gap between the conduction band
and the valence band which we not show here.
From Fig.7(b) and (c), we can see that the larger the Hubbard U is, the wider the JDOS-curve expand in the low-energy region.
While in the positive-energy-region, the most dominantly curve is the one in critical-U and gradually decrese when away from the critical-U.
(see the distribution of antibonding states in Fig.7(c) in the range of $E>10$ eV).
Thus the scattering strength is reaches the maximum at the point of critical U.
In these DOS plots, we also found that the $p$-band is dominate no matter how large the U is,
except for the low-energy region.
For the pure monolayer silicene and 2nd AA-stacked bilayer silicene (Fig.7(d) and (e), respectively),
the critical Hubbard U for the transition from semimetal phase to insulator is 9 eV and 12 eV, respectively.

We next do a dynamical analysis for the in-plane ac conductivity 
of monolayer silicene.
Here we comment that for the 1st AA-stacked bilayer one, 
the Hall conductivity of the clear sample (without impurity) shows differences with the monolayer one, like the appearence of the $\sigma_{xy}=0$ plateau,
which disapear in the monolayer due to the gapless Dirac-coone when without the electric field or exchange field 
but with the effective SOC and Rashba-coupling (see Ref.\cite{Wu C H}).
The $\sigma_{xy}=0$ plateau is appears in the case that the chemical potential 
$|\mu|<{\rm min}\{t_{{\rm inter}}\}$ where $t_{{\rm inter}}$ is the interlayer hopping\cite{Hsu Y F},
since the gapped characteristic as shown in the band structure of Fig.2.
The in-plane optical conductivity of monolayer silicene with a finite impurity strength, 
$\sigma(\Omega)=\sigma_{xx}+\sigma_{xy}$ is shown in the Fig.8,
which with the peaks center around 5 eV.
The origin of these peaks is associated with VHS in the DOS\cite{Mousavi H}.
Here we note that the momentum change can be evaluated as $\Delta{\bf k}=\hbar(\Omega+i\delta)$ in the analytic continuation
with the bosonic frequency (photon) $\Omega$.
The above critical value of Hubbard U which close to 12 eV is also valid for the in-plane longitudinal conductivity as we shown in the figure.
From Fig.8, we find that there is a large decrease of the value of conductivity in the critical-U both for the real part and the imaginay part.
That also exhibit a behavior that transfer to the band insulator phase, where we use a dash-line to devides the two parts (semimetal phase and insulator phase).

This critical-U is also far away from that of the Kane-Mele-Hubbard (KMH) model phase diagram which is U=4.3$t$ 
measured by the method of Quantum Monte Carlo (QMC) as we did in the Ref.\cite{Wu C H}.
In the KMH model for the hexagonal lattice at half-filling, the semimetal phase transfer to the AFM Mott insulator at the critical U=4.3$t$
for the case of only consider the NN hopping.
A analogous value obtained by QMC which is 4.5$t$ is reported in Refs.\cite{Santoro G,Sorella S}.
But the common points with our present model is that
the phase transition of nonmagnetic semimetal phase to the AFM Mott insulator will becomes
the paramagnetic phase to the AFM Mott insulator when consider the long-range hopping (like the $t'$),
and the strength of SDW would be reduced compared to the $d_{1}+id_{2}$ SC.
While for the phase transition from the AFM order Mott insulator to the semimetal can be realized by increasing the strength of exchange field
to a value comparable with $\lambda_{SOC}$\cite{Ezawa M},
in this process the AFM superexchange enhance the correlations between the singlet pairing and may results in the valence band resonance (RVB),
which can also be realized by the manipulation of the impurities AFM correlation by the impurity doping.
The AFM correlation term of the impurities here can be written as $\sum_{\langle i,j \rangle}J_{ij}(S_{i}S_{j}-1/4)$\cite{Baskaran G},
where the superexchange factor $J_{ij}=4t^{2}/U$ in the AFM region is estimeted as in the range of 0.85 to 1.49
in our model.
The optical conductivity below the critical value also become numb with the change of U unlike the ones above the critical value.
After the quenching of the kinetic energy in the microcosmic process due to the momentum transfer,
in equilibrium stage, the peak is the mid-infrared peak local at the frequency as twice of the
exchange scale\cite{Jaklič J}.
It's also clearly that the peaks of optical conductivity is shift rightwards with the increase of U, which also
exhibits the decrease of the strength of electron-phonon interaction,
i.e., the electron-phonon interaction red-shift the absorption features of optical conductivity spectra\cite{Havener R W}.
With the increase of on-site Hubbard U,
the kinetic energy also decrease with the increasing quasiparticle mass.

Using the random-phase-approximation (RPA) approach, the above-mentioned long-range Hubbard repulsion 
(NN Hubbard repulsion)
could be screened by the high energy state which with large charge DOS (like the VHS).
The screened in-plane long-range Hubbard repulsion can be written as 
\begin{equation} 
\begin{aligned}
g_{s}({\bf k},\omega)=\frac{g}{1-4g\Pi({\bf k},\omega)},
\end{aligned}
\end{equation}
where $g$ is the universe Coulomb repulsion as $g=2\pi e^{2}/(\epsilon_{0}{\bf k})$.
The factor 4 here denotes the number of degenerate (or the flavors).
Here the wave vector ${\bf k}$ denotes the position in the momentum space which not restricted around the Dirac-point,
and it can be replaced by $v_{F}$ to obtain the dimensionless long-range form\cite{Khveshchenko D V} $g_{0}$.
Here we comment that with the increase of this dimensionless Coulomb repulsion $g_{0}$,
the linear relation of DOS in low-energy tends to quadratic relation\cite{Guinea F2}.
The denominator of the above expression can be view as the non-static dielectric function $\epsilon^{-1}({\bf k},\omega)$ obtained by RPA,
where the energy loss function $L$ can be well obtained through the relation $L={\rm Im}\epsilon^{-1}({\bf k},\omega)$ as shown in Ref.\cite{Wu C H}.
Or in the optical language,
the above dielectric function can be rewritten as $\epsilon^{-1}({\bf k},\omega)=1-\frac{8\pi^{2}c_{s}}{{\bf k}}\Pi({\bf k},\omega)$,
where $c_{s}$ is
the Sommerfeld vacuum fine structure constant
$c_{s}=\frac{e^{2}}{2\epsilon_{0}h c}=1/137.036$\cite{Nair R R}.
which is related to the zero-$\omega$ optical absorption in the limit of vanishing SOC by $A_{op}(0)=\pi c_{s}$\cite{Matthes L2},
and is applicable for all of the group IV atoms.
The complex polariztion function (or the susceptibility) $\Pi({\bf k},\omega)$ 
can be deduced from the retarded current-current correlation function in bubble diagram as\cite{Gusynin V P,Nicol E J}
\begin{equation} 
\begin{aligned}
\Pi({\bf k},\Omega)=-\frac{4e^{2}}{\beta}\int\frac{d^{2}k}{4\pi^{2}}{\rm Tr}[v_{\alpha}G_{{\bf k}}(i\omega+\Omega+i\delta)v_{\beta}G_{{\bf k}}(i\omega)],
\end{aligned}
\end{equation}
where $G_{{\bf k}}(i\omega)=\int^{\infty}_{-\infty}\frac{d\omega}{2\pi}\frac{A(\omega,{\bf k})}{i\omega+\mu-\omega}$
with $A(\omega,{\bf k})$ the spectral weight.
$v_{\alpha}$ and $v_{\beta}$ denote the two velocity operators with the leads $\alpha,\beta=x,y,z$,
which
\begin{equation} 
\begin{aligned}
v_{x}=v_{F}I\gamma_{x},\ v_{y}=v_{F}I\gamma_{y},\ v_{z}=v_{F}I\gamma_{z},
\end{aligned}
\end{equation}
with $I$ the $4\times 4$ identity matrix,
and the $4\times 4$ Gamma matrices: $\gamma_{x}=\sigma_{z}\otimes i\sigma_{y}$, 
$\gamma_{y}=\sigma_{z}\otimes i\sigma_{x}$, $\gamma_{z}=i\sigma_{z}\otimes i\sigma_{z}$.
While for the screened interlayer Hubbard repulsion which is rised by the interlayer interaction,
is\cite{González J,Baskaran G2}
\begin{equation} 
\begin{aligned}
g'_{s}({\bf k},\omega)=\frac{g\ {\rm sinh}(d{\bf k})}{\sqrt{({\rm cosh}(d{\bf k})+g\ {\rm sinh}(d{\bf k})\Pi({\bf k},\omega))^{2}-1}},
\end{aligned}
\end{equation}
with the above polarization function can be rewritten as 
\begin{equation} 
\begin{aligned}
\Pi({\bf k},\Omega)=-\frac{4}{S}\sum_{\Delta{\bf k}}\frac{f_{{\bf k}+\Delta{\bf k}}-f_{{\bf k}}}{E_{{\bf k}+\Delta{\bf k}}-E_{{\bf k}}-\Omega-i\delta}.
\end{aligned}
\end{equation}
The polariztion function will becomes $\omega$-independent for the interband transition\cite{Hwang E H} between the conduction band and valence band
which only happen in the strong Coulomb-coupling case in the monolayer silicene\cite{Wu C H}.

In conclusion, we investigate the manipulation of the phase transition of the semimetal silicene to the Mott insulator
(or the paramagnetic Mott insulator in the present of long-range hopping)
without applying the magnetic field or the laser beam.
There are not nesting at the zero filling for the silicene (see the map-plot in Fig.4),
and the AFM order is absent at the begining with U=0.
That provides the premise for the phase transition from semimetal to insulator.
The AFM Mott insulator phase can emerges under the U larger that the critical value expressly for the AB-stacked bilayer silicene
or the multilayer bulk one\cite{Harigaya K},
althought the silicene is not bipartite like the graphene due to its strong intrinsic SOC
except viewed as a composite of two opposite triangular sublattices\cite{Sahin H}.
In the presence of the on-site Hubbard U in our tight-binding model with the induced layer potential difference,
the competing with the spin-dependent exchange field $M_{s}$ may leads to the topological phase transition 
between zero and nonzero Chern number\cite{Wu C H},
and with the tunable Hall current under the bias energy.
The charged impurity is proved that affects deeply the DOS of the monolayer and bilayer silicene in this paper,
and induced the long-range Coulomb scattering which with the 
mean elastic diffusion distance $\frac{v_{F}}{2\Gamma}\sim\sqrt{n}$ in the zero-temperature limit.
It's also been found that the increasing of impurity concentration may reduce the critical temperature for the phase transition\cite{Iye Y}.
and the phase transition to the paramagnetic Mott insulator has been proved to be second order in the decrease of double occupation\cite{Ebrahimkhas M}.
The increasing Hubbard U also leads to the renormalization of the kinetic energy or the Fermi velocity in the presence of electron interactions,
and even the renormalization quasiparticle mass.
We also find that the triplet excitons in the Mott insulator region may arised with the increasing $d_{1}+id_{2}$ pairing instability
due to the decresing of long-range hopping (like the $t'$) which may reduce the inter-sublattice symmetry. 
The RVB which with the observable charge fluctuation also arised near the critical value for the phase transition of semimetal to insulator
(or the paramagnetic silicene to insulator for the case with nonzero NNN hopping).
Through the charge impurity scattering together with the Coulomb repulsion, 
the transport properties (like the optical conductivity) are explored with the different short-range Hubbard interaction in this paper.
Finally, in contrast to the AB-stacked bilayer silicene or the multilayer bulk one,
the monolayer silicene or theAA-stacked bilayer one have weaker AFM or FM (excitonic) instability and the exchange instability
under the large on-site Hubbard U exceeds the critical value.
That also related to the 
SC $d_{1}+id_{2}$ type wave pairing which can be emerged in the bilayer silicene and
has the common characteristics of the $d$-wave SC: like the anisotropic dispersion which also affects the transoprt properties of the charge carriers,
and the enhanced particle-hole pairing (like the Cooper pair) strength.

\section{Acknowledgement}
I thank Peng-Cheng Li for useful discussions.

\end{large}
\renewcommand\refname{References}

\clearpage
Fig.1
\begin{figure}[!ht]
\subfigure{
\begin{minipage}[t]{0.5\textwidth}
\centering
\includegraphics[width=1.1\linewidth]{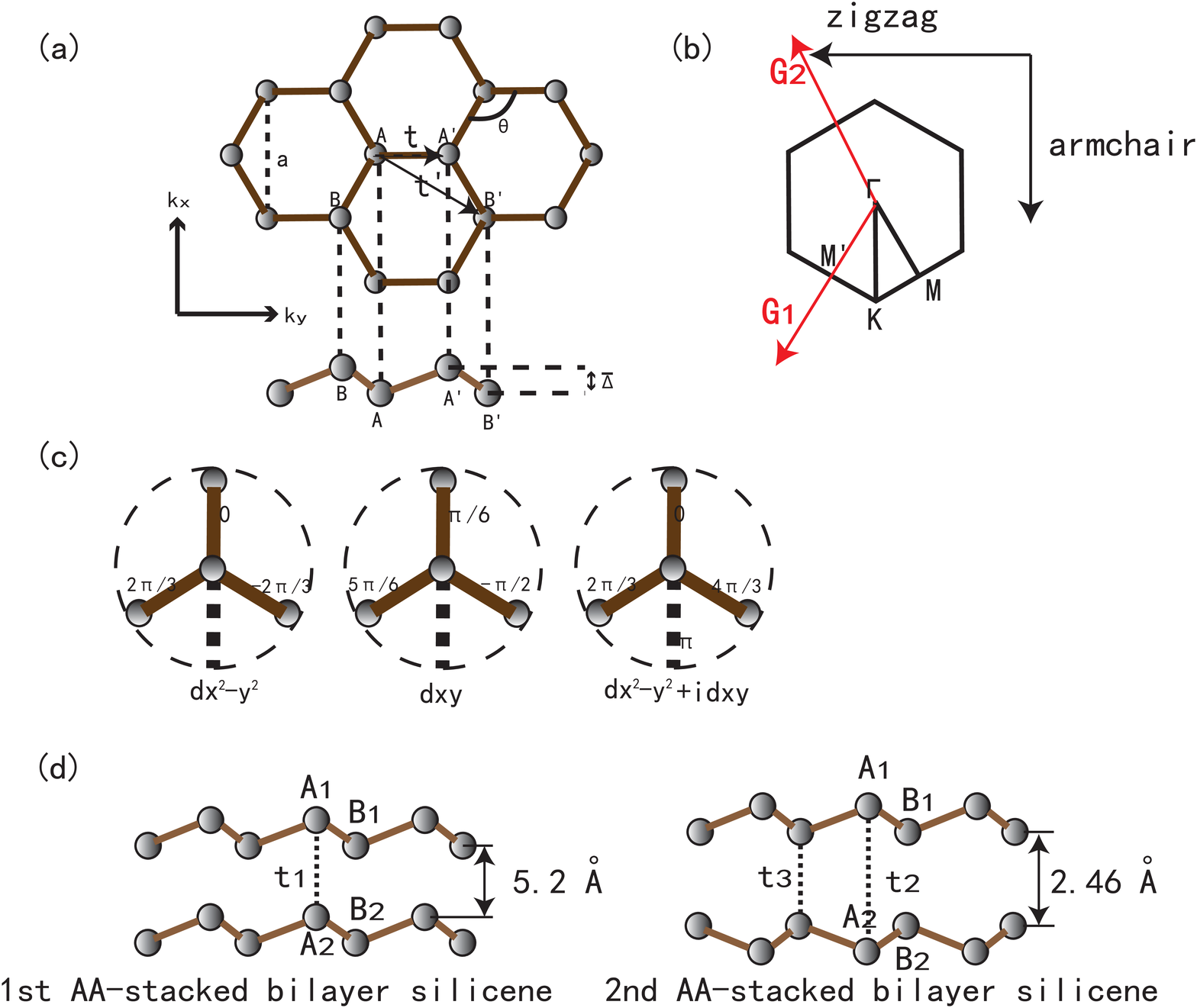}
\label{fig:side:a}
\end{minipage}
}
\subfigure{
\begin{minipage}[t]{0.5\textwidth}
\centering
\includegraphics[width=0.9\linewidth]{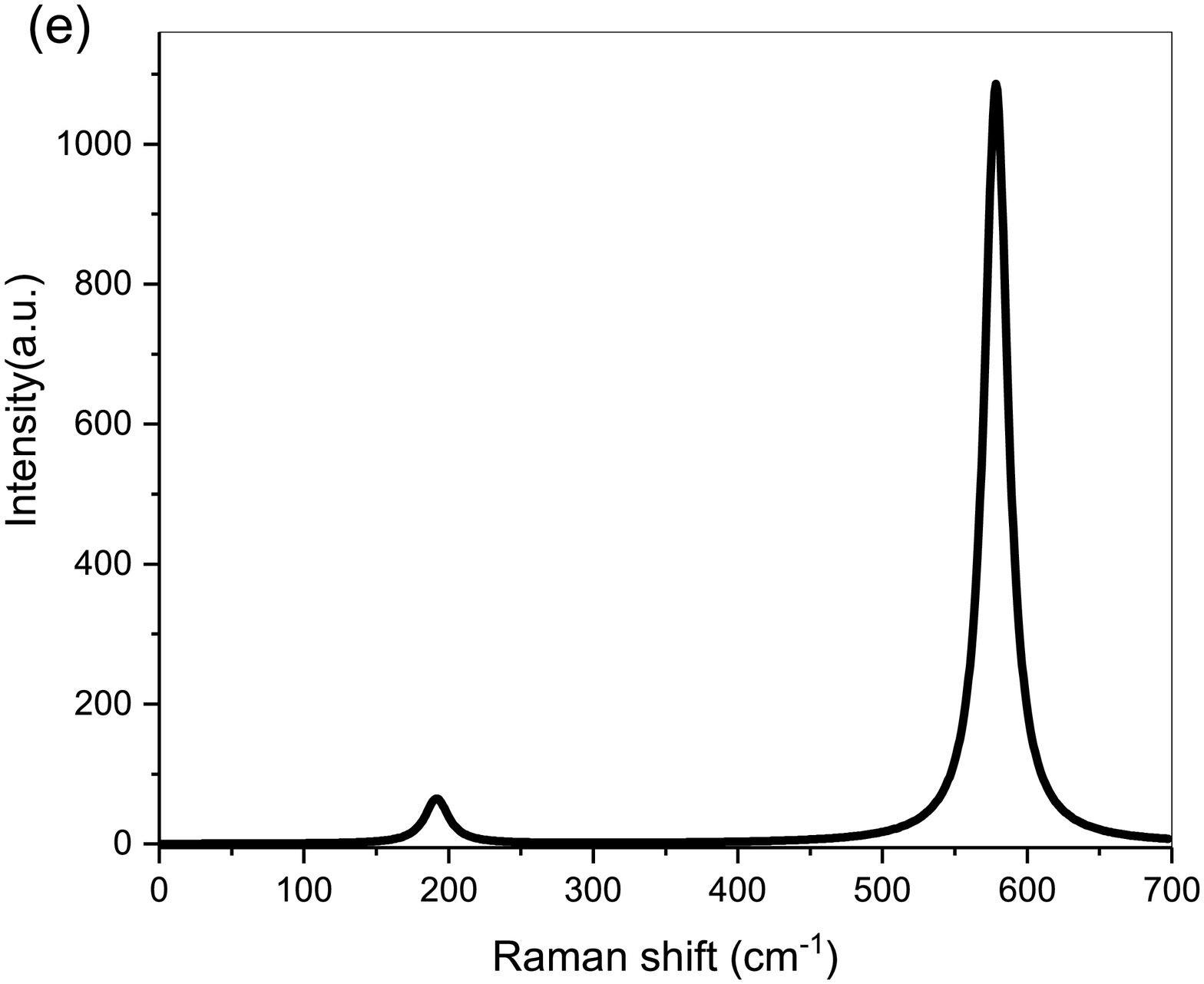}
\label{fig:side:b}
\end{minipage}
}
\caption{(a)Top view and side view of the silicene.
with four sites (sublattices) $A,B,A',B'$ in unit cell. 
 The bond-angle $\theta$ and the buckling distance $\Delta$ were marked.
The three dashed lines with $t,t',t''$ denotes the
nearest-, second nearest-, and third nearest-neighbor hopping, respectively.
The blue and green solid lines denotes the hopping in $r$ direction and $r'$ direction respectively,
where $r'$ contains the three hopping directions which goven by the phase $\phi$ and $r$ contains the three ones which not goven by the phase $\phi$.
(b) Brillouin zone (the $k$-space) with the high symmetry points.
The Red vector in the right panel is the reciprocal lattice vector ${\bf G}_{1}=(\frac{-2\sqrt{3}\pi}{3a},-\frac{2\pi}{a}),
{\bf G}_{2}=(\frac{-2\sqrt{3}\pi}{3a},\frac{2\pi}{a})$.
(c) Phase of the $d_{x^{2}-y^{2}},\ d_{xy},\ d_{x^{2}-y^{2}}+id_{xy}$ pairing symmetries (left to right) of silicene in real space.
(d) the two kinds of the AA-stacked silicene: the first one
with the nearest layer distance as 5.2 \AA\ and intra-layer bond length 2.28 \AA\ and with bulked distance $\overline\Delta=0.46$ \AA\ the same as the monolayer one,
the second one with the nearest layer distance as 2.46 \AA\ and intra-layer bond length 2.32 \AA\ and with lattice constant $a=3.88$,
and the bulked distance becomes $\overline\Delta=0.64$ \AA\.
The interlayer hopping label in the figure are $t_{3}\simeq2$ eV $>t_{2}>t_{3}$.
(e) The Lorentzian fit of the Raman spectrum of monolayer silicene.
}
\end{figure}

\clearpage
Fig.2
\begin{figure}[!ht]
\subfigure{
\begin{minipage}[t]{0.15\textwidth}
\includegraphics[width=1\linewidth]{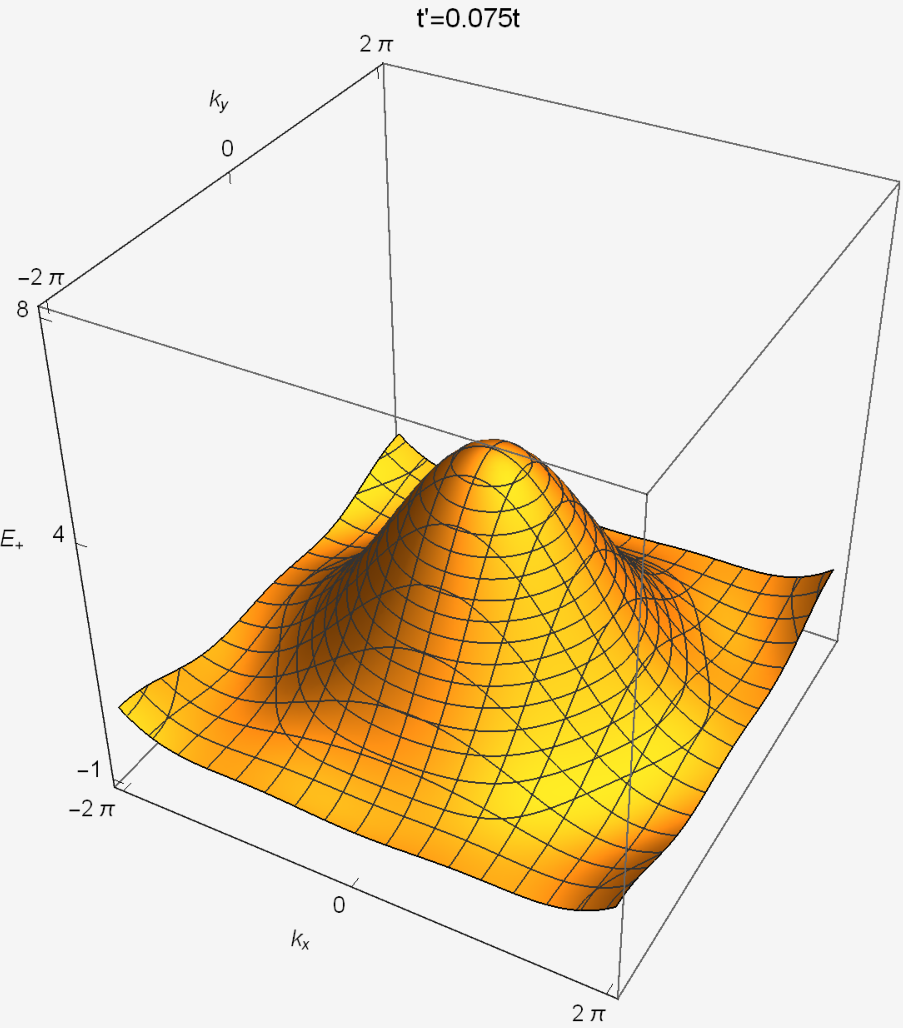}
\label{fig:side:a}
\end{minipage}
}
\subfigure{
\begin{minipage}[t]{0.15\textwidth}
\includegraphics[width=1\linewidth]{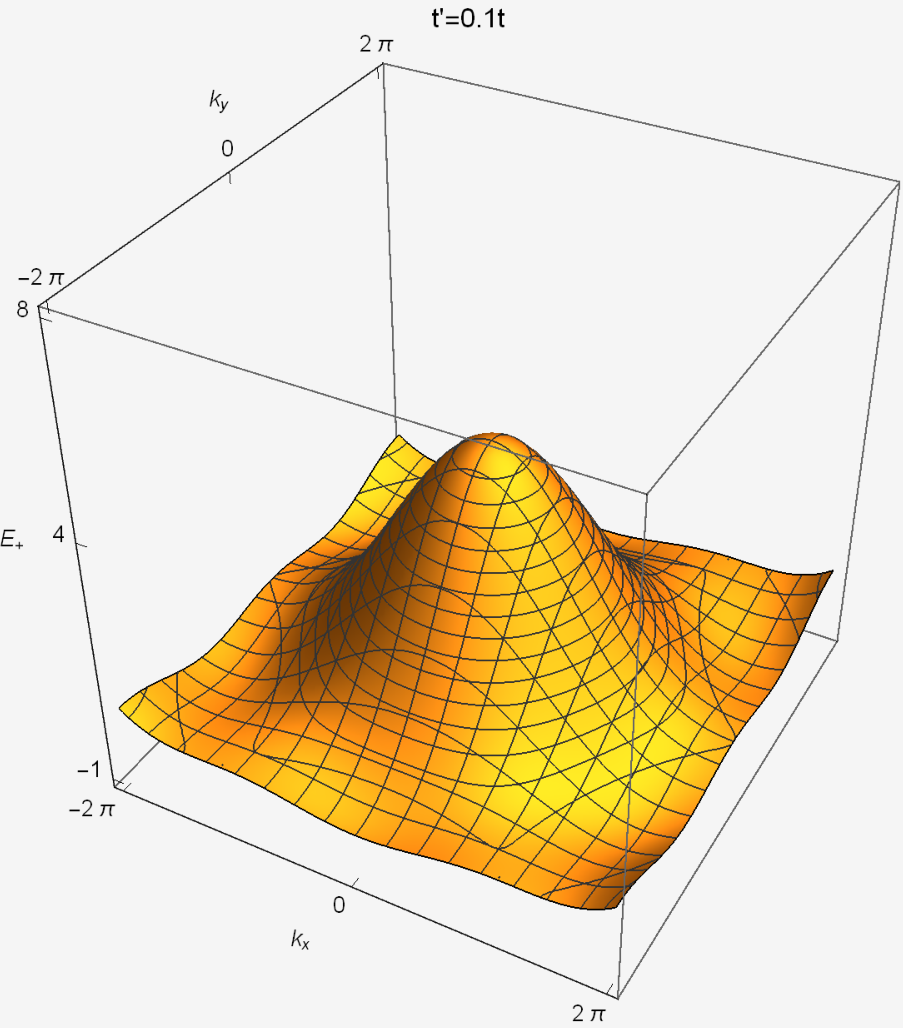}
\label{fig:side:b}
\end{minipage}
}
\subfigure{
\begin{minipage}[t]{0.15\textwidth}
\includegraphics[width=1\linewidth]{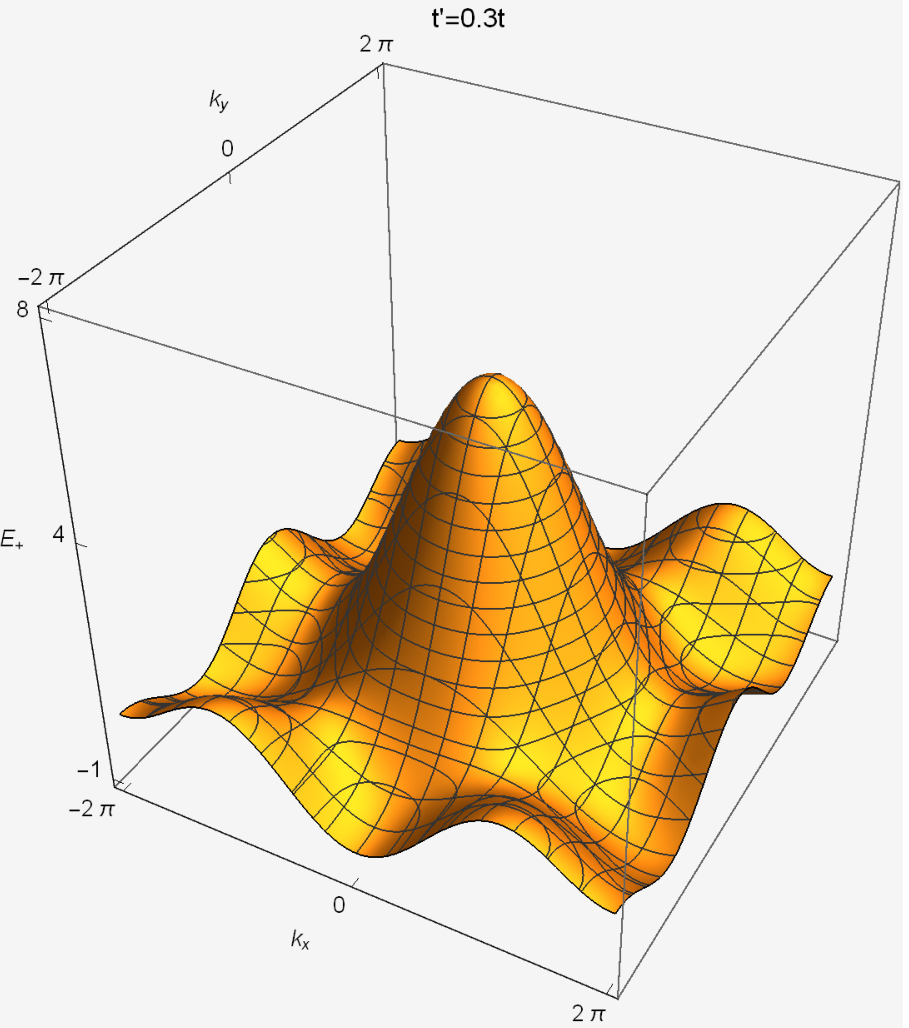}
\label{fig:side:a}
\end{minipage}
}
\subfigure{
\begin{minipage}[t]{0.15\textwidth}
\includegraphics[width=1\linewidth]{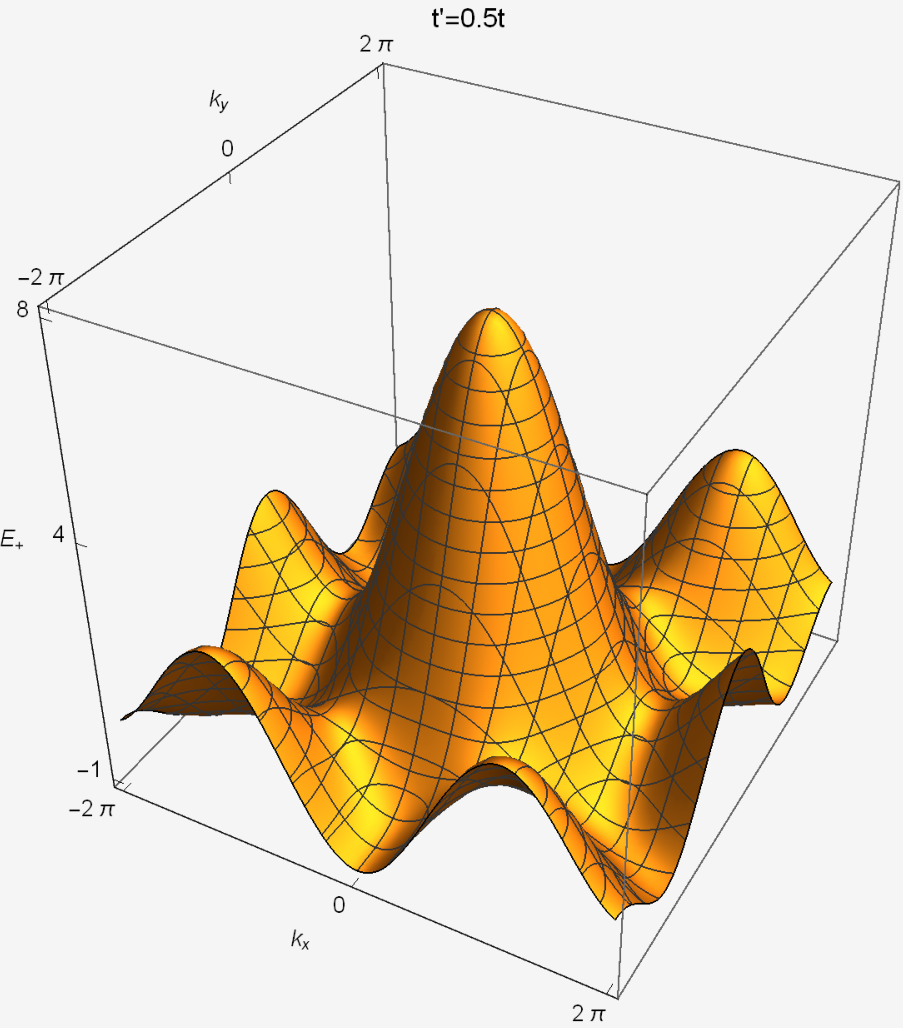}
\label{fig:side:b}
\end{minipage}
}\\
\subfigure{
\begin{minipage}[t]{0.15\textwidth}
\includegraphics[width=1\linewidth]{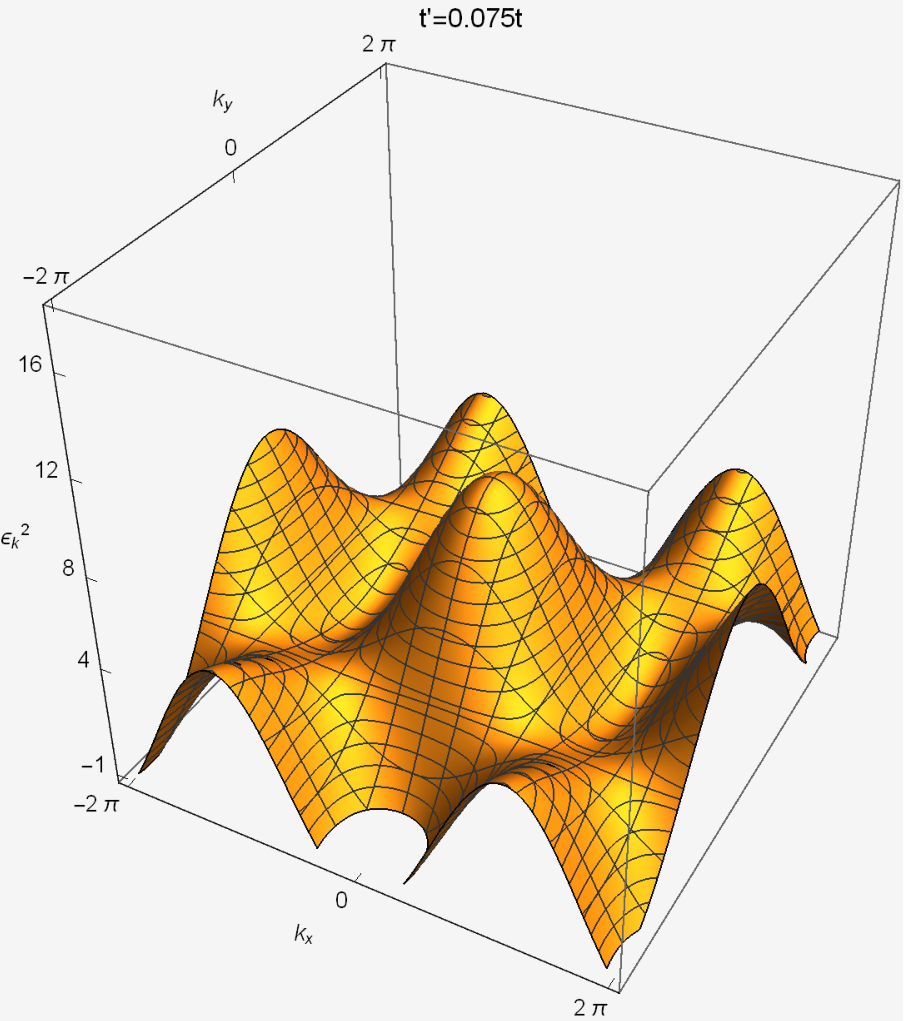}
\label{fig:side:a}
\end{minipage}
}
\subfigure{
\begin{minipage}[t]{0.15\textwidth}
\includegraphics[width=1\linewidth]{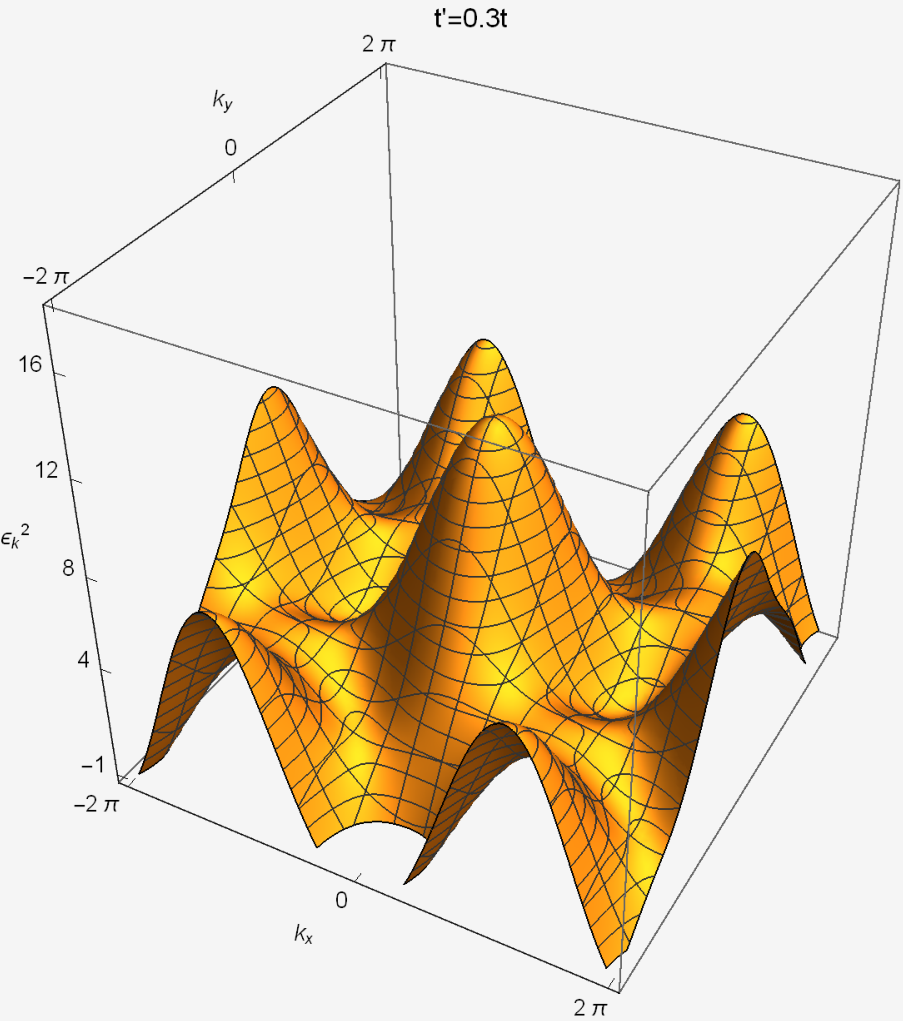}
\label{fig:side:b}
\end{minipage}
}
\subfigure{
\begin{minipage}[t]{0.15\textwidth}
\includegraphics[width=1\linewidth]{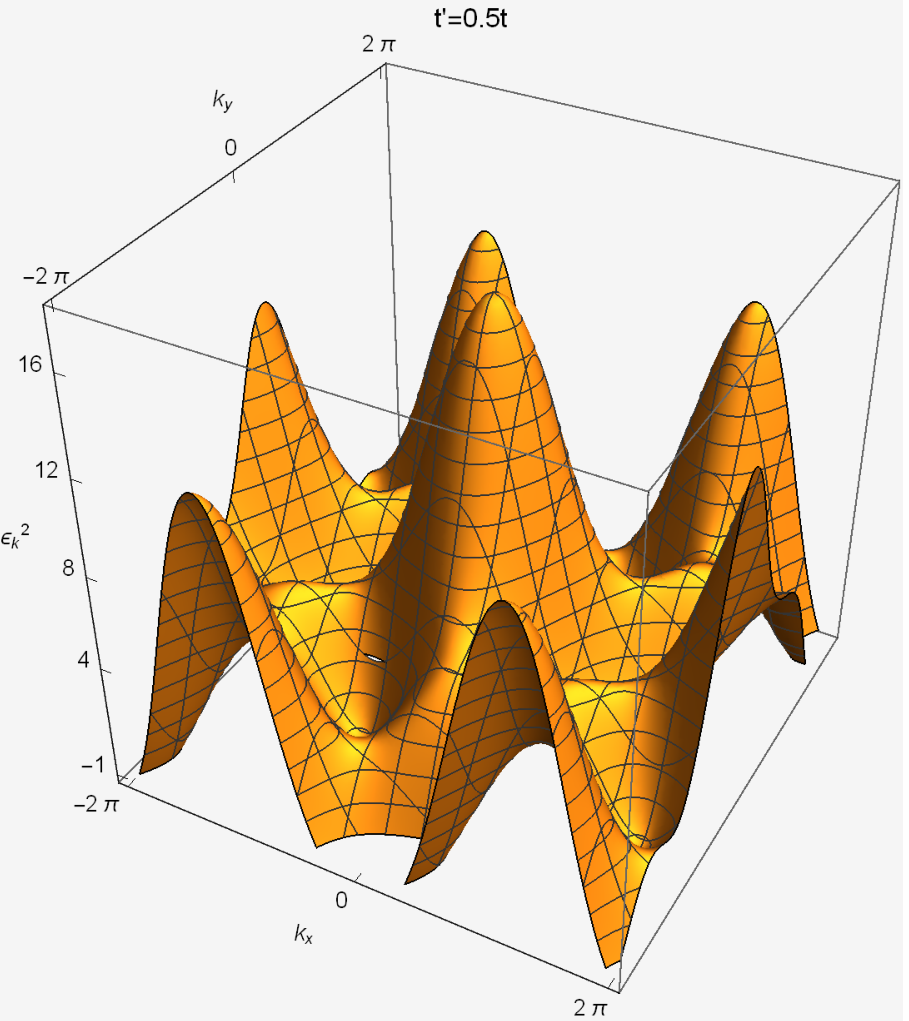}
\label{fig:side:a}
\end{minipage}
}
\subfigure{
\begin{minipage}[t]{0.15\textwidth}
\includegraphics[width=1\linewidth]{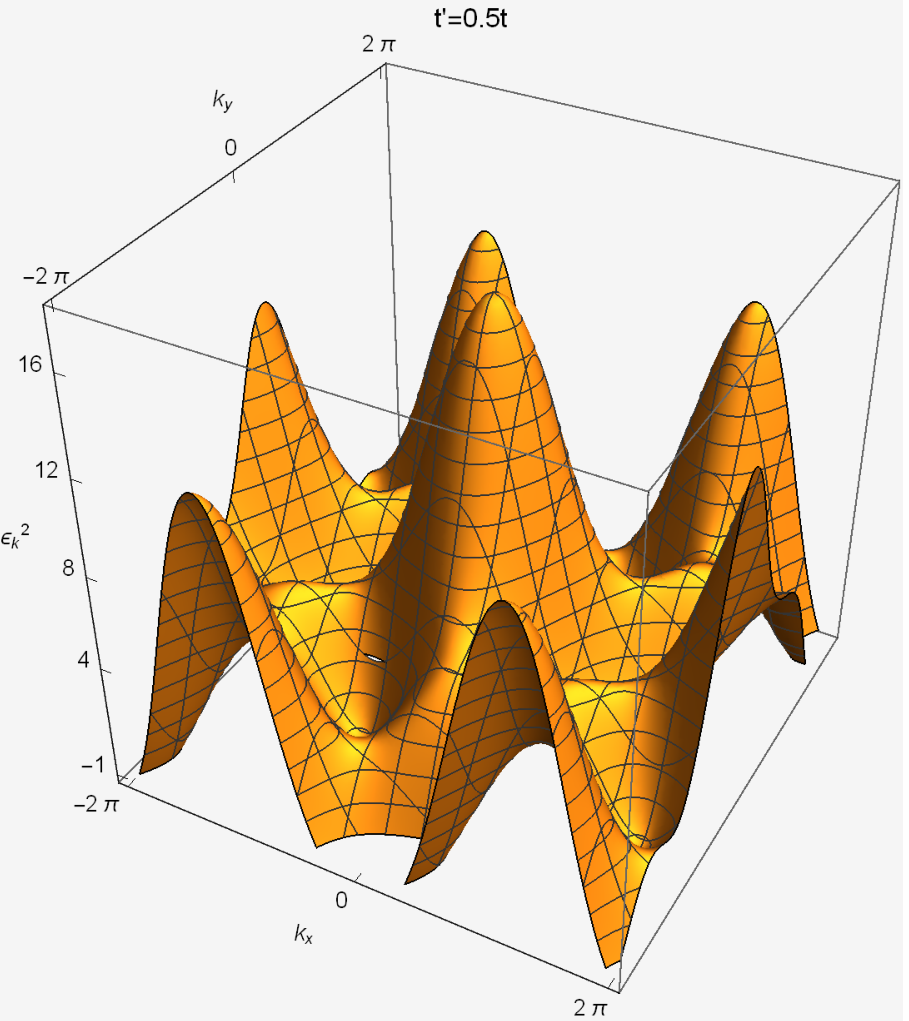}
\label{fig:side:b}
\end{minipage}
}
\caption{(Color online) 3D Schematic diagram of the band structures of monolayerd silicene in momentum space
obtained from Eq.(2) and Eq.(3), respectively.
The upper panel is the upper bands energy and lower panel the energy in the single-particle picture with different $t'$.
The on-site energy is setted as 1 here and the NN hopping is setted as 1 for simplify.
The NNN hopping are 
$t'=0.075t,\ t'=0.1t,\ t'=0.3t,\ t'=0.5t$ from left to right.}
\end{figure}

\clearpage
Fig.3
\begin{figure}[!ht]
   \centering
   \begin{center}
     \includegraphics*[width=0.5\linewidth]{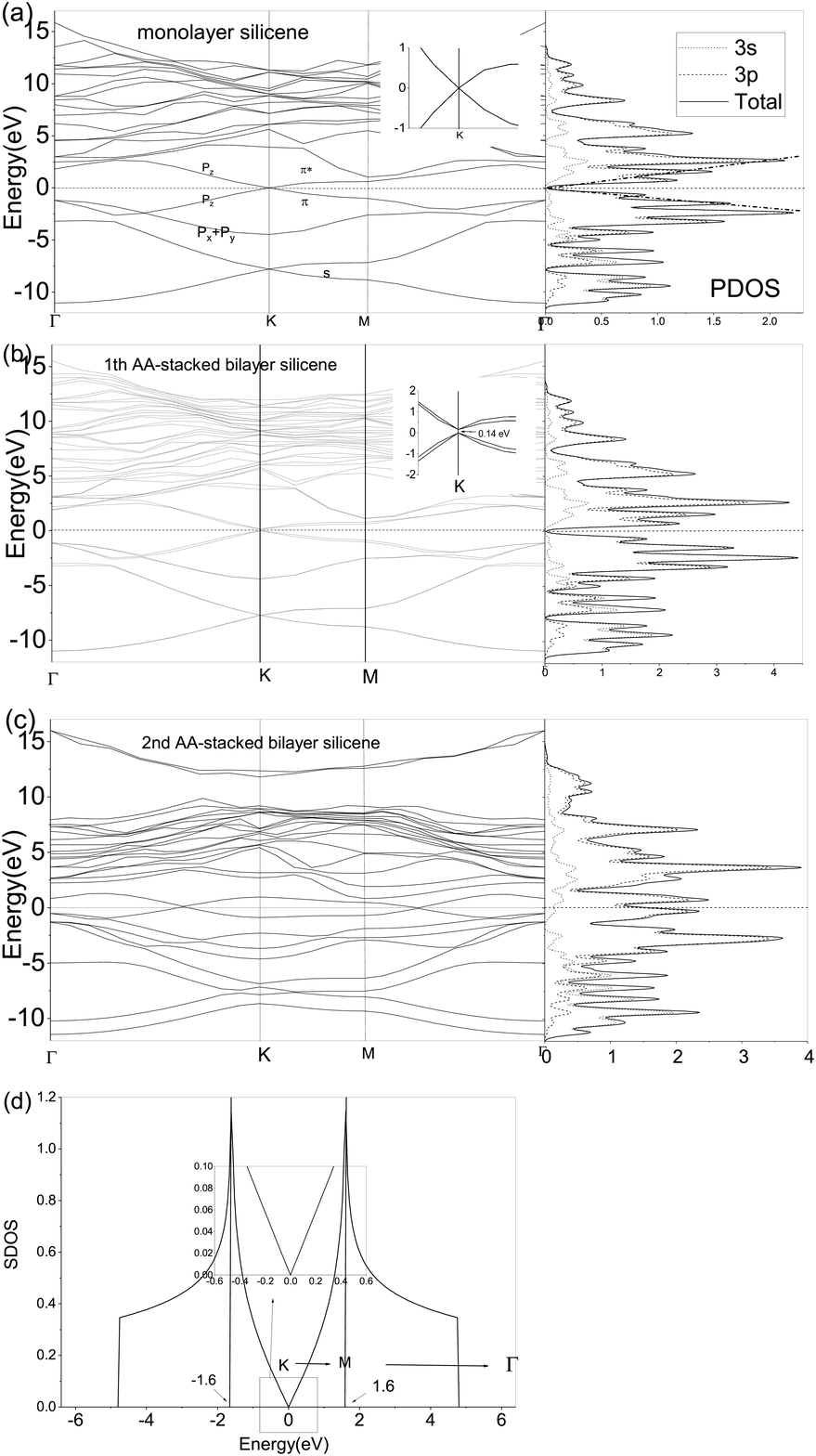}
\caption{Band structure of the monolayer silicene (a), 1st AA-stacked bilayer silicene (b), 
and 2nd AA-stacked bilayer silicene (c) as well as their PDOS in the right side.
The insets show the enlarged views of band structure in the K-point,
which for the bilayer 1st AA-stacked one exhibits a gap as 0.14 eV in the K-point (and the gap 0.15 eV is obtained by using the method of
local-density approximation (LDA)), while it's gapless for the monolayer one.
There's a slight difference for our results about the 1st AA-stacked silicene from the Ref.\cite{Liu F}s',
which exhibit a band clossing as 0.2 eV much smaller than that of the bilayer graphene which is $2t_{{\rm inter}}$\cite{Tabert C J}.
That may due to the different inter-layer separations of the sample,
e.g., see Ref.\cite{Kamal C}.
In (c), the band stucture of 2nd AA-stacked silicene exhibit two cross-point in the range of $\Gamma-K$ and $\Gamma-M$, respectively.
The $\pi^{*}$-band and $\pi$-band which are cross in the Dirac-point are labeled in (a),
and the bands main contributed by $p_{z}$, $p_{x}+p_{y}$ , and $s$ orbit are also labeled.
From the PDOS, we can clearly see that the silicene is a $3p$-orbital-based materials,
but the 3s-orbit is dominate below the -5 eV due to its large electronegativity.
The PDOS in (a) is similar to the figure of the single-particle DOS (d) which consider only the NN-hopping here.
The two Van-Hove singularities emerge at $\varepsilon=\pm 1.6$ eV,
which corresponds to the hexagon Fermi surface (nexted) enclosed by the six M-point (the saddle points of the band structure in first BZ
) of the as labeled in the figure (see also the Fig.2),
while for the case with impurities or the lattice defects, the zero-energy point (neutrality point)
exhibits a smeared $\delta$-function with a finite width\cite{Bena C,He J}),
They both has a linear dispersion near the neutrality point (see the inset of (c) and the dash-dot line in the right-side of (a)).
}
   \end{center}
\end{figure}

\clearpage

Fig.4
caption:(Color online) Map plots (equal value contours) of the tight-binding energy dispersion (upper panel) 
and their corresponding DOS (bottom panel) for the silicene with particle-hole symmetry
where we ignore the broken of inversion symmetry by the bulked structure and the Rashba-coupling (NNN) and consider only the NN hopping.
The DOS-map are obtained by the renormalization group method in momentum space.
In (a) and (b), we consider the dispersion in hexagonal BZ with the contribution from $t=1.6$ eV and $t=V_{pp\pi}^{(1)}$
(which measured as -0.72 in Ref.\cite{Guzmán-Verri G G} and -1.12 in Ref.\cite{Liu C C}) respectively,
while in (c) we consider the dispersion in square BZ with $t=1.6$ eV and with Dirac mass $m_{D}=$0.32 eV. 
The distribution of the DOS is the same as the one shown in Fig.1(d),
i.e., arrives the maximum value in the hexagonal Fermi surface where the Fermi surfce is nested now.
For the case of $t=-0.72$, we can't find the nested hexagonal Fermi surface anymore.
Note that we only takes the real part for the computing results.

\clearpage

Fig.5
\begin{figure}[!ht]
   \centering
   \begin{center}
     \includegraphics*[width=0.8\linewidth]{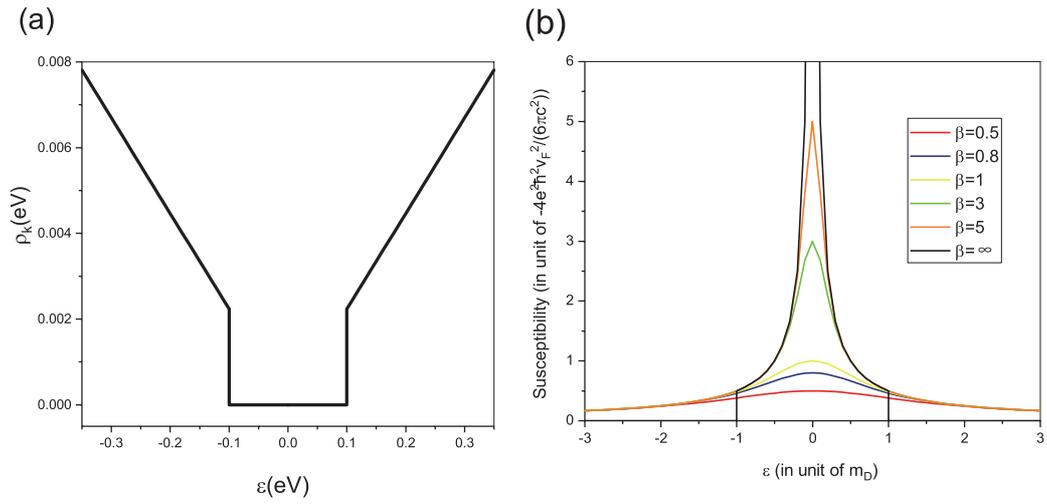}
\caption{(Color online)
(a) DOS at zero temperature as a function of the energy where we set $|\lambda_{SOC}+M|=0.1\ {\rm eV},R=0,E_{\perp}=0$.
(b)
The negative orbital susceptibility as a function of the energy in unit of $m_{D}$
under the zero-temperature limit $\beta\rightarrow\infty$ and a series of finite temperatures.}
   \end{center}
\end{figure}

\clearpage

Fig.6
\begin{figure}[!ht]
   \centering
   \begin{center}
     \includegraphics*[width=0.8\linewidth]{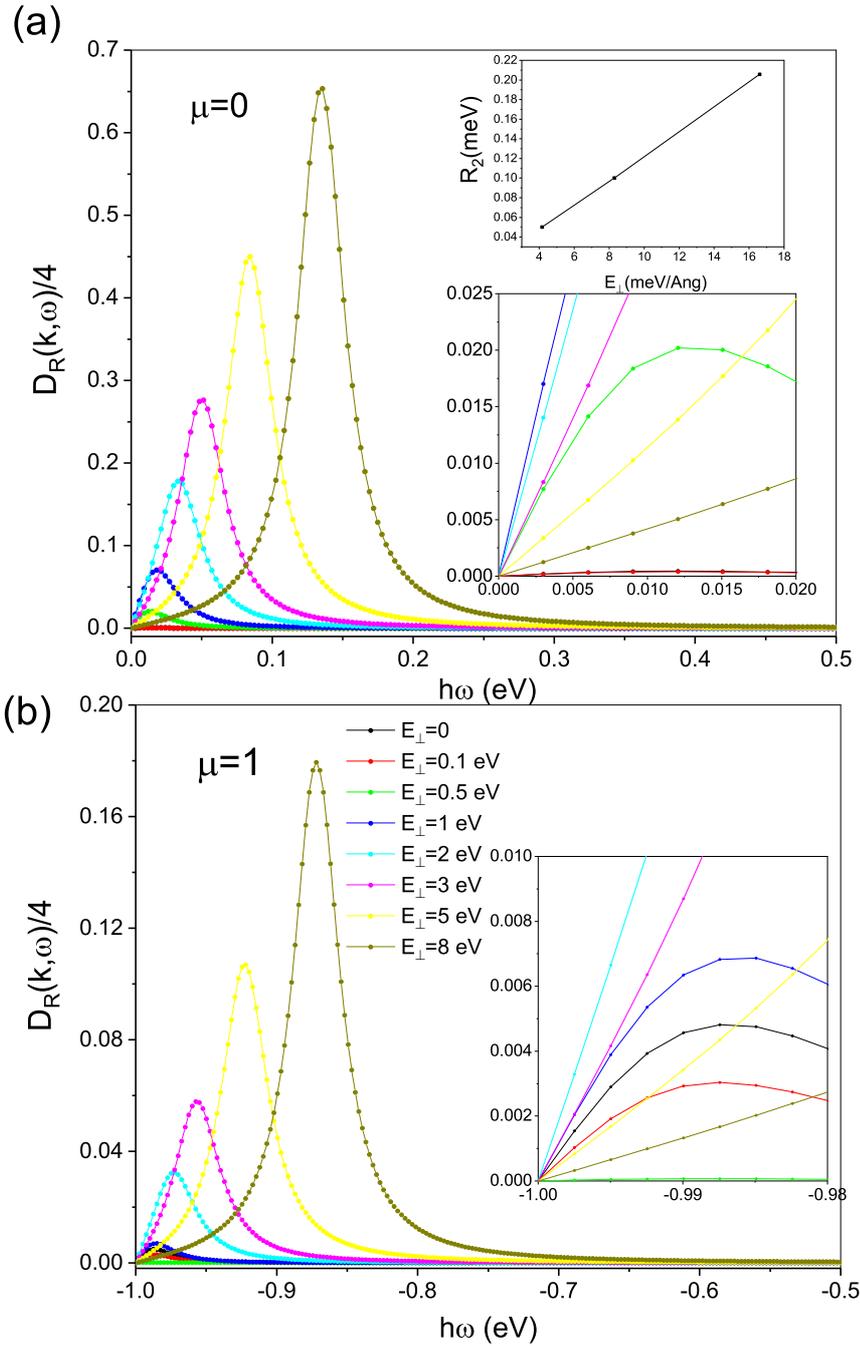}
\caption{(Color online) The DOS contributed by the Rashba-coupling under different perpendicular electric field strength. 
The temperature $k_{B}T$ is setted as 0.025 eV and the chemical potential is zero (a) and 1 (b), respectively.
When the electric field is zero, there exist only the NNN intrinsic Rashba-couping; when the electric field was applied,
the induced NN Rashba-coupling $R_{2}(E_{\perp})$ is follow the linear relation with the strength of the applied electric field, $R_{2}(E_{\perp})=0.012E_{\perp}$,
as shown in the upper inset.
Note that we simplify the $\Gamma$ as 0.01 in the computational process of this figure.}
   \end{center}
\end{figure}

\clearpage
Fig.7
\begin{figure}[!ht]
\subfigure{
\begin{minipage}[t]{0.5\textwidth}
\centering
\includegraphics[width=1.1\linewidth]{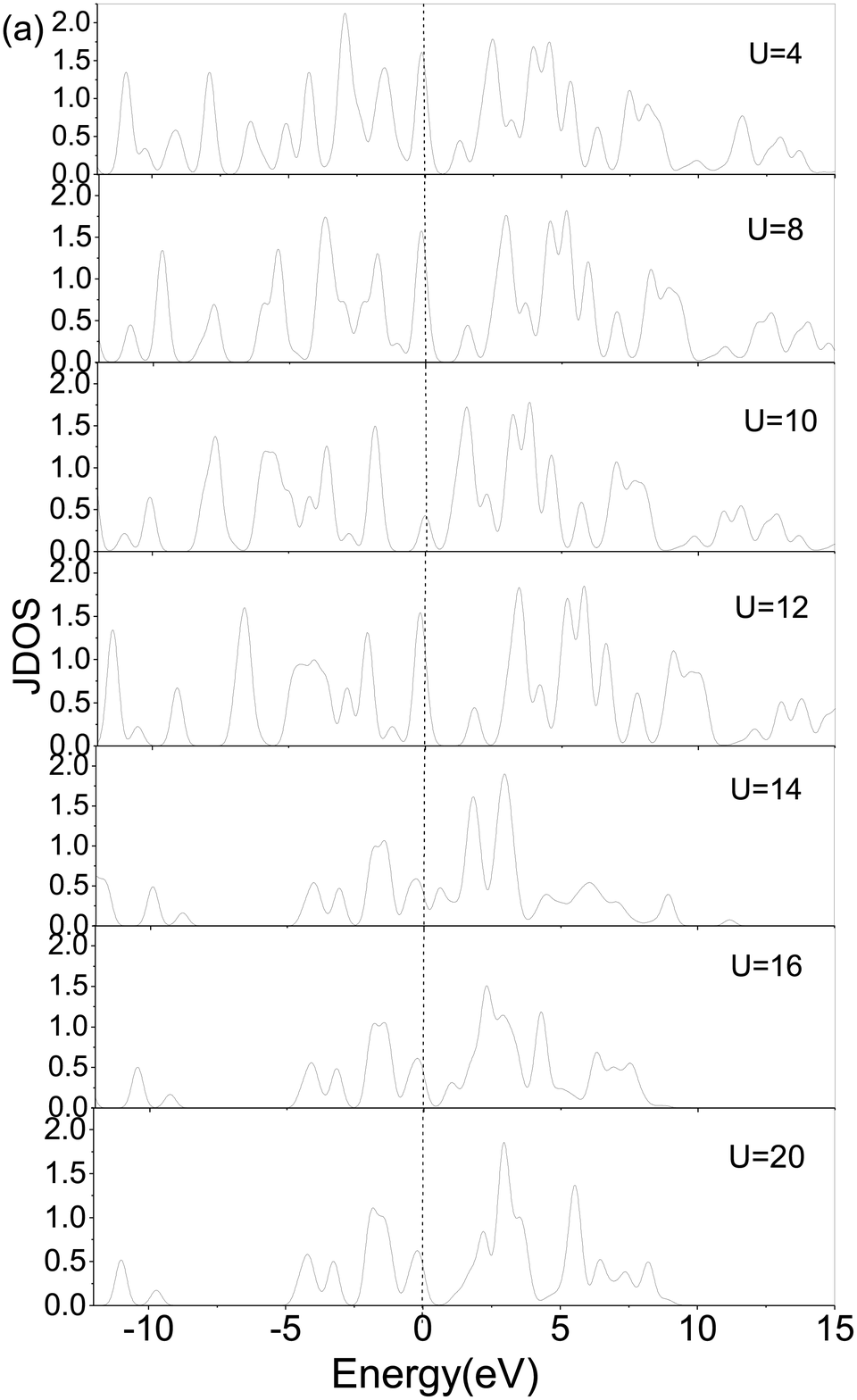}
\label{fig:side:a}
\end{minipage}
}
\subfigure{
\begin{minipage}[t]{0.5\textwidth}
\centering
\includegraphics[width=0.9\linewidth]{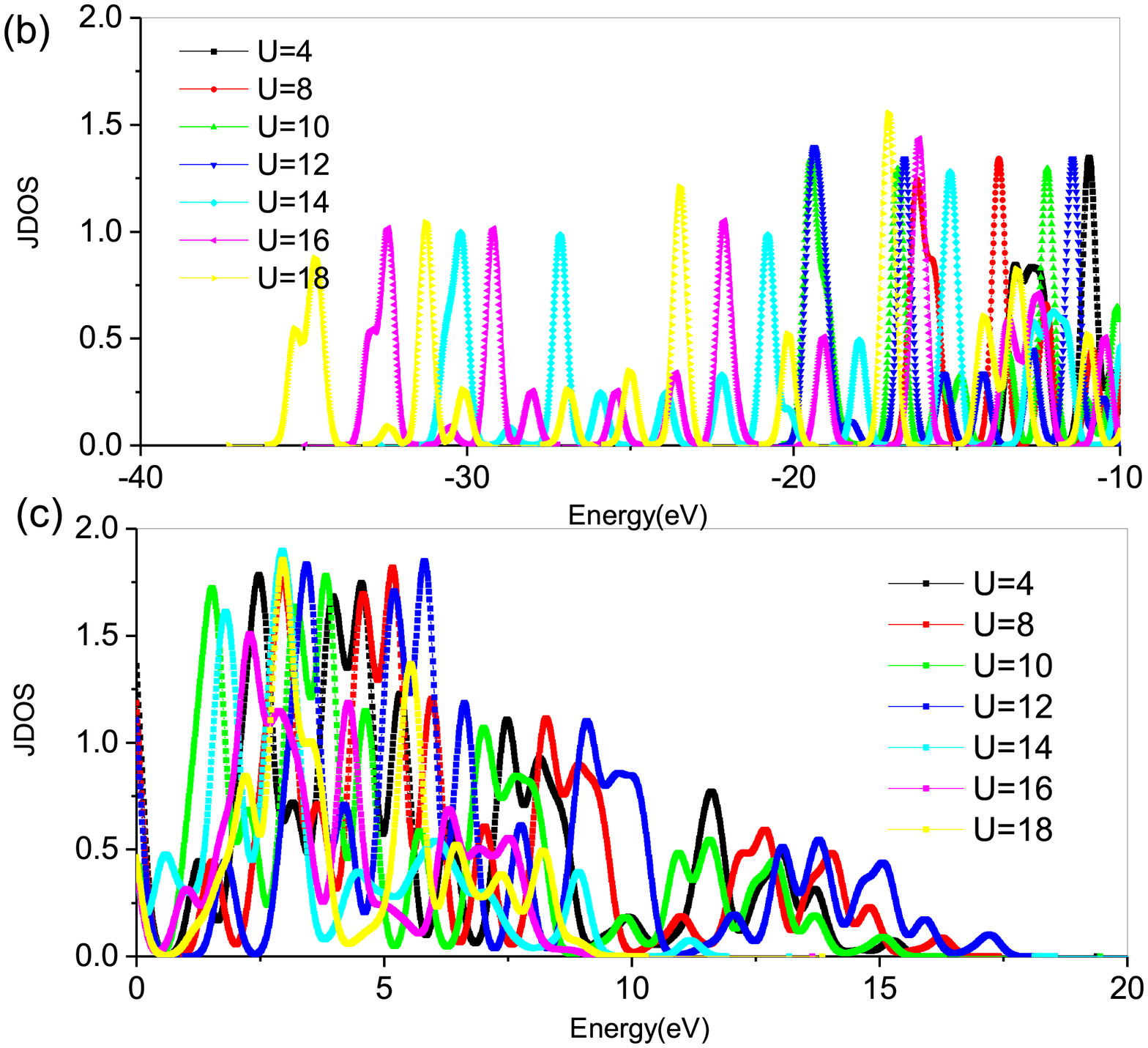}
\label{fig:side:b}
\end{minipage}
}
\subfigure{
\begin{minipage}[t]{0.5\textwidth}
\centering
\includegraphics[width=0.5\linewidth]{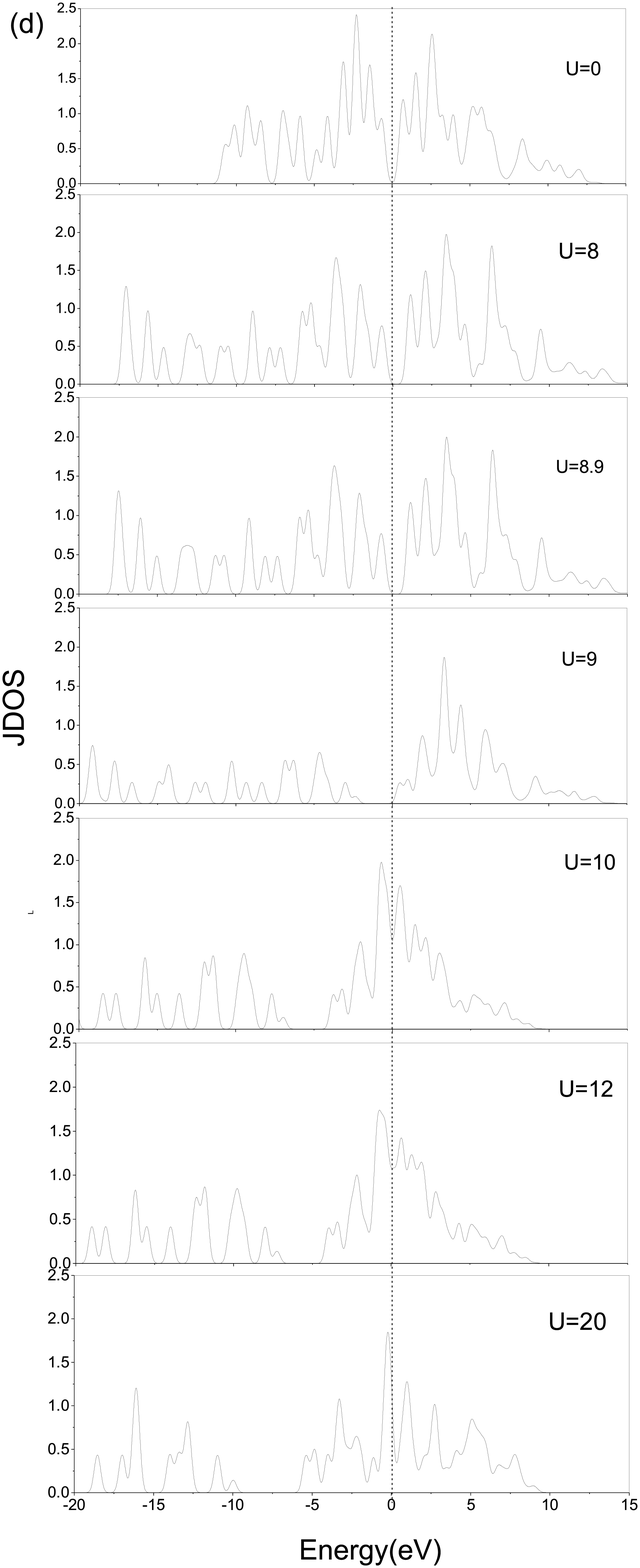}
\label{fig:side:b}
\end{minipage}
}
\subfigure{
\begin{minipage}[t]{0.4\textwidth}
\centering
\includegraphics[width=0.5\linewidth]{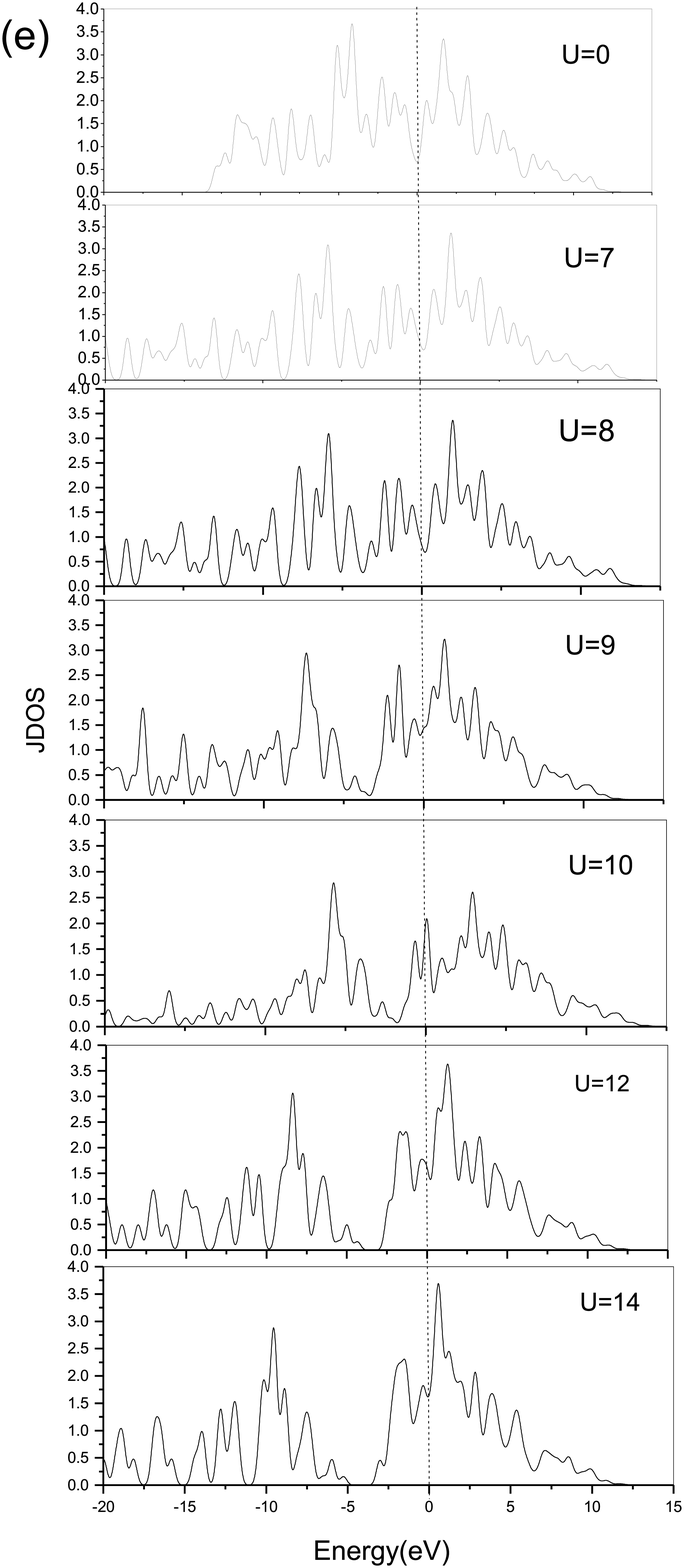}
\label{fig:side:b}
\end{minipage}
}
\caption{(Color online)(a) Joint DOS (JDOS) of monolayer silicene with the impurity whose strngth is 4 eV under different Hubbard U. 
The value of Hubbard U are 
labeled in the plot.
The zero-energy level are labeled by the dot-line.
Note that the silicene here is in AFM order.
(b) and (c) shows the JDOS (a) in positive-energy-region and negative-energy-region, respectively.
For the pure monolayer silicene (d) and the pure 2nd AA-stacked bilayer silicene (e), 
we show the resulting JDOS around the critical U.
}
\end{figure}

\clearpage
Fig.8
\begin{figure}[!ht]
   \centering
   \begin{center}
     \includegraphics*[width=0.8\linewidth]{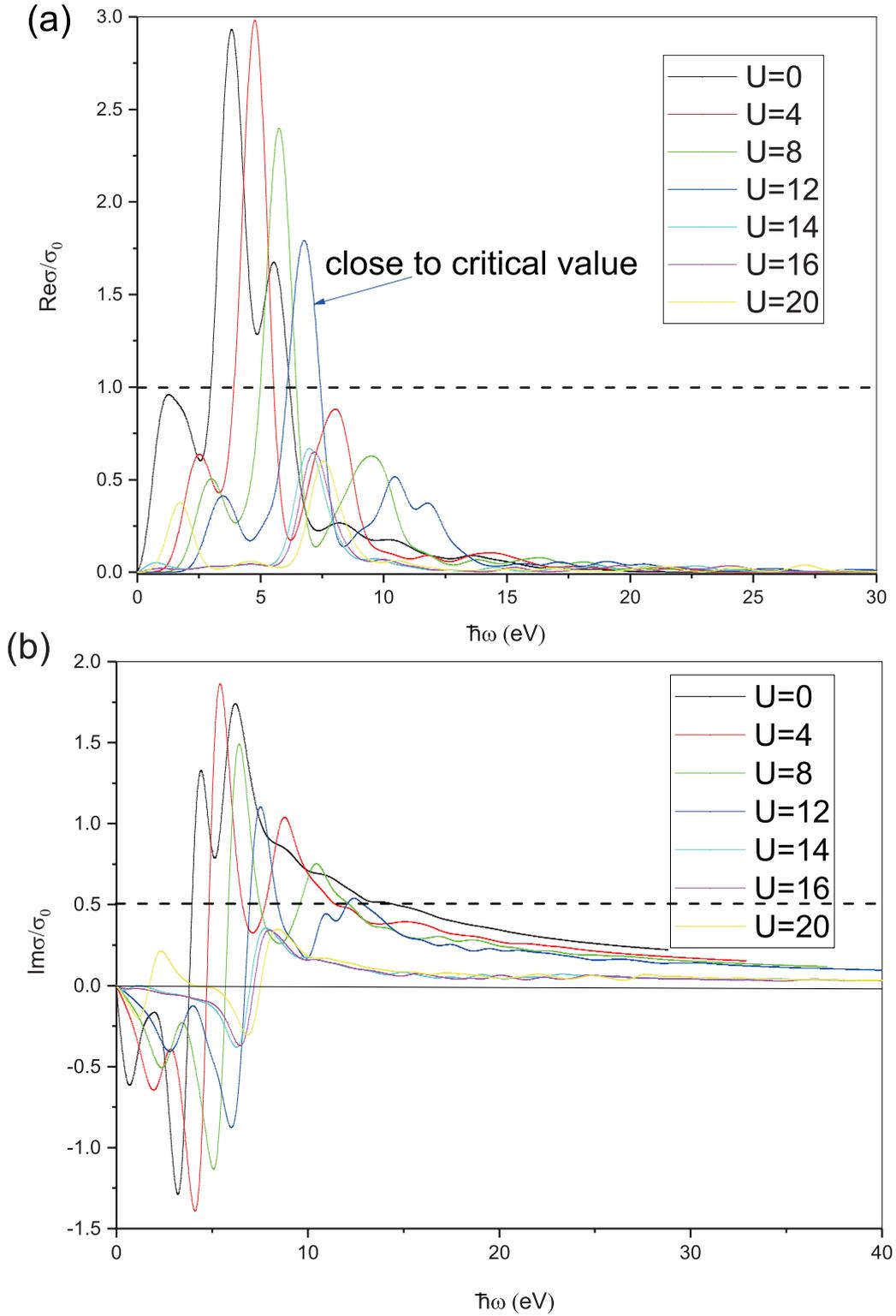}
\caption{(Color online) Real part (a) and imaginary part (b) of the optical conductivity of the dirty monolayer silicene with the impurity strength 4 eV
(corresponds to Fig.7(a)-(c))
under different Hubbard U. The vertical-axis is in unit of ac constant conductivity $\sigma_{0}=e^{2}/(4\hbar)$,
which is valid for all the group-IV monolayer honeycomb crystals, and $\sigma_{0}=e^{2}/(2\hbar)$ for the bilayer one.
}
   \end{center}
\end{figure}

\end{document}